\documentclass[pra,aps,showpacs,twocolumn,superscriptaddress]{revtex4}
\usepackage{bbm}
\usepackage{amsmath,amsfonts,amssymb,mathrsfs,bm}
\usepackage{graphicx,graphics,color,epsfig}
\begin{document}
\title{Binary mixture of pseudospin-$\frac{1}{2}$ Bose gases with interspecies spin exchange: from classical fixed points and ground states to quantum ground states}
\author{Rukuan Wu}
\affiliation{State Key Laboratory of Surface Physics and  Department
of Physics, Fudan University, Shanghai 200433, China}
\affiliation{Department of Physics, Zhejiang Normal University,
Jinhua 321004, China}
\author{Yu Shi}
\email{yushi@fudan.edu.cn}
\affiliation{State Key Laboratory of Surface Physics and  Department
of Physics, Fudan University, Shanghai 200433, China}

\begin{abstract}

We consider the effective spin Hamiltonian describing a mixture of two species of pseudo-spin-$\frac{1}{2}$ Bose gases with
interspecies spin exchange. First we analyze the stability of the fixed points of the corresponding classical
dynamics, of which the signature is found in quantum dynamics with a disentangled initial state.
Focusing on the case without an external potential, we find all the ground states by taking into account quantum fluctuations around the classical ground state in each parameter regime. The nature of entanglement and its relation with classical bifurcation is investigated. When the total spins of the two species are unequal, the maximal entanglement at the parameter point of classical bifurcation is possessed by the excited state corresponding to the classical fixed point which bifurcates, rather than by the ground state.

\end{abstract}


\maketitle

\section{Introduction}

A remarkable discovery in recent years is that bifurcation in classical dynamics is related to quantum entanglement in the ground state of the corresponding quantum Hamiltonian. In addition to its theoretical demonstration in the Dicke model~\cite{milburn1,bif5}, in a model of two coupled giant spins describing magnetic clusters~\cite{bif1}, and in an integrable
dimer model \cite{bif2}, this association has also been studied in the area of Bose-Einstein condensation  (BEC), including
two-component  BEC \cite{bif3a,bif3} and two-mode atomic-molecular  BEC \cite{bif4,bif4a}.  More recently,  a classical bifurcation has been
observed in an experiment realizing an internal Josephson effect
in a spinor Bose-Einstein condensate, as an important step toward
entanglement generation close to critical points~\cite{bif6}.
Moreover, in a laser-cooled atom, experimental evidence has been observed for entanglement being a quantum signature of chaos~\cite{chaos}.

On the other hand, a novel kind of BEC, the so-called EBEC, that is, BEC of interspecies spin singlet pairs, was found to be the ground state of a Bose system composed of two species of pseudo-spin-$\frac{1}{2}$
Bose atoms with both intraspecies and interspecies spin-exchange
interactions in a considerable parameter regime \cite{shi0,shi,shi2,shi3,wu}. Under the usual single orbital-mode approximation, the Hamiltonian of this system can be transformed into that of two coupled giant spins. Alas, the ground states in all parameter regimes have not yet worked out.

In this paper, we make  each of the above two lines of research useful for the other. We focus on the case in the absence of an external potential.
First, we study the bifurcation of the classical dynamics corresponding to the Hamiltonian of this Bose mixture, by analyzing the
stability of each  fixed point. Quantum dynamics displays some features similar  to the classical dynamics if the initial state is a disentangled state, which, however, is not an energy eigenstate.   When the numbers of the atoms of the two species are equal, a bifurcation of the fixed points indeed corresponds to a
quantum phase transition to a maximally entangled ground
state. When they are unequal, the quantum state corresponding to the classical fixed point which bifurcates is also maximally entangled at the bifurcation point. However, it is not the ground state. Finally, we analytically obtain all the quantum ground states by considering quantum fluctuations around the classical ground state in each parameter regime. The analytical results fit the numerical results very well.

The rest of this paper is organized as follows. The model is introduced in Sec.~\ref{model}. The fixed points and bifurcations are studied in Sec.~\ref{bifurcation}, with the detailed calculation reported in the Appendix. The classical ground states are described in Sec.~\ref{classical}, and the classical evolution is described in Sec.~\ref{evolution}, whose quantum analog is described in Sec.~\ref{quantum}. Section ~\ref{entanglement} describes the absence  of the correspondence between classical bifurcation and maximal entanglement  in the case of unequal populations of the two species. In Sec.~\ref{ground}, we find the quantum ground state in each parameter regime by approximating the Hamiltonian around the classical ground state there. Finally the paper is summarized in Sec.~\ref{summary}.

\section{The Model \label{model}}

Consider a dilute gas composed of two distinct species of Bose atoms with the following property~\cite{shi,shi2}.
Each atom has an internal degree of freedom represented as a
pseudospin-$\frac{1}{2}$, with $z$-component basis states $\uparrow$
and $\downarrow$. Under the usual single orbital-mode approximation, for
each  species $\alpha (\alpha = a,b)$ and pseudospin
$\sigma$($\sigma=\uparrow,\downarrow$), only the single-particle
orbital ground state $\phi_{\alpha\sigma}(\textbf{r})$ is occupied,
then the Hamiltonian can be transformed into that of two coupled
giant spins with spin quantum numbers $S_a=N_a/2$ and $S_b=N_b/2$. Here we focus on the uniform case~\cite{shi3,wu},  for which
\begin{equation}
{\cal
\hat{H}}=J_{\perp}(\hat{S}_{ax}\hat{S}_{bx}+\hat{S}_{ay}\hat{S}_{by})+
J_z \hat{S}_{az}\hat{S}_{bz} \label{spinhamiltonian}
\end{equation}
where $J_{\perp}=4\pi \hbar^2\xi_e/(m_{ab}\Omega)$ while $J_z=4\pi \hbar^2(\xi_s-\xi_d)/(m_{ab}\Omega)$,  with $m_{ab}$ being the reduced mass of an $a$ atom and a $b$ atom,  $\xi_e$ is the scattering length for the scattering in which an $a$ atom and a $b$ atom exchange pseudospins,  $\xi_s$ being the scattering length for the forward  scattering in which an $a$ atom and a $b$ atom have different pseudospins without spin exchange,
$\xi_d$ being the scattering length for the forward  scattering in which an $a$ atom and a $b$ atom have the same pseudospin without spin exchange,  $\Omega$ being the volume of the system, $\mathbf{\hat{S}}_{\alpha}=
\hat{\alpha}^\dagger_\sigma\mathbf{s}_{\sigma\sigma'}\hat{\alpha}_{\sigma'}$,
$\mathbf{s}_{\sigma\sigma'}$ is the single spin operator,
$\alpha_\sigma$ denotes the annihilation operator associated with
$\phi_{\alpha\sigma}(\textbf{r})$ of species $\alpha$. Without loss of generality, suppose $S_a \geq S_b$.

The corresponding classical Hamiltonian is obtained from (\ref{spinhamiltonian}) by treating the spin operators as the classical spin variables,
\begin{eqnarray}
{\cal  H}_{cl} &=&J_{\perp}(S_{ax}S_{bx}+S_{ay}S_{by})+
J_z S_{az}S_{bz} \label{spinhamiltonian2}\\
&= &J_\perp\sqrt{(S_a^2-S_{az}^2)(S_b^2-S_{bz}^2)}
\cos(\varphi_a-\varphi_b) \nonumber \\
&& +J_zS_{az}S_{bz} \label{spinhamiltonian3}
\end{eqnarray}

From the Hamiltonian (\ref{spinhamiltonian}), one obtains the equations of motion
\begin{equation}
\begin{array}{rl}
\frac{d\hat{S}_{\alpha x }}{dt} & = J_\perp \hat{S}_{\alpha z}\hat{S}_{\beta y}-J_z \hat{S}_{\alpha y}\hat{S}_{\beta z},\\
\frac{d\hat{S}_{\alpha y}}{dt} & =-J_\perp \hat{S}_{\alpha z}\hat{S}_{\beta x}+J_z \hat{S}_{\alpha x}\hat{S}_{\beta z},\\
\frac{d\hat{S}_{\alpha z}}{dt} & = J_\perp(\hat{S}_{\alpha
y}\hat{S}_{\beta x}-\hat{S}_{\alpha x}\hat{S}_{\beta y}).
\label{motion1}
\end{array}
\end{equation}

The corresponding classical equations of motion, obtained either from the classical Hamiltonian (\ref{spinhamiltonian2}) or from the quantum equations of motion by  replacing the spin operators as the classical spin variables, can be written as
\begin{equation}
\frac{d \mathbf{A}}{dt} = {\cal J} \mathbf{A}, \label{motionclassical}
\end{equation}
where $\mathbf{A} \equiv (S_{ax},S_{ay},S_{az},S_{bx},S_{by},S_{bz})^T$,
\begin{widetext}
$$ {\cal J} \equiv \left(
  \begin{array}{cccccc}
    0 & -J_z S_{bz} & J_\perp S_{by} & 0 & J_\perp S_{az} & -J_z S_{ay} \\
    J_z S_{bz} & 0 & -J_\perp S_{bx} & -J_\perp S_{az} & 0 & J_zS_{ax} \\
    -J_\perp S_{by} & J_\perp S_{bx} & 0 & J_\perp S_{ay} & -J_\perp S_{ax} & 0 \\
    0 & J_\perp S_{bz} & -J_zS_{by} & 0 & -J_zS_{az} & J_\perp S_{ay} \\
    -J_\perp S_{bz} & 0 & J_z S_{bx} & J_z S_{az} & 0 & -J_\perp S_{ax} \\
    J_\perp S_{by} & -J_\perp S_{bx} & 0 & -J_\perp S_{ay} & J_\perp S_{ax} & 0 \\
  \end{array}
\right). $$
\end{widetext}
In studying the stability of a fixed point, ${\cal J}$ becomes the Jacobian matrix when the spin variables adopt the values at this fixed point.

The classical Hamiltonian in the form of (\ref{spinhamiltonian3}) suggests that the classical
state of the system is completely determined  by the variables
$S_{az}$, $S_{bz}$, and $\varphi_a-\varphi_b$.

We shall use (\ref{spinhamiltonian3}) studying the evolution in Sec.~\ref{evolution}, while using (\ref{spinhamiltonian2}) in analyzing the fixed points in Sec.~\ref{bifurcation}, because there is arbitrariness of angles $\phi_{a}$ and $\phi_b$ in some fixed points.

\section{Fixed points and bifurcations in classical dynamics \label{bifurcation} }

The fixed points of the classical dynamics are obtained from
\begin{equation}
\frac{d\mathbf{A}}{dt}=0. \label{fpequation}
\end{equation}

The stability of each fixed point can be examined first by studying
the eigenvalues of Jacobian matrix ${\cal J}$ at this point: It is stable if every eigenvalue has a negative real part, while it is unstable if any eigenvalue has a positive real part. Otherwise, one cannot judge whether the fixed point is
stable from the eigenvalues of ${\cal J}$, but it is certainly stable if a
Lyapunov function can be constructed.
A Lyapunov function ${\cal F}$ is such that in a
neighborhood of the fixed point, ${\cal L}$ is minimal (or maximal)
at the fixed point, and $d{\cal L}/dt \leq 0$(or $d{\cal L}/dt \geq
0$).

\begin{figure}
\begin{center}
\includegraphics[height=2.5in,width=3.0in]{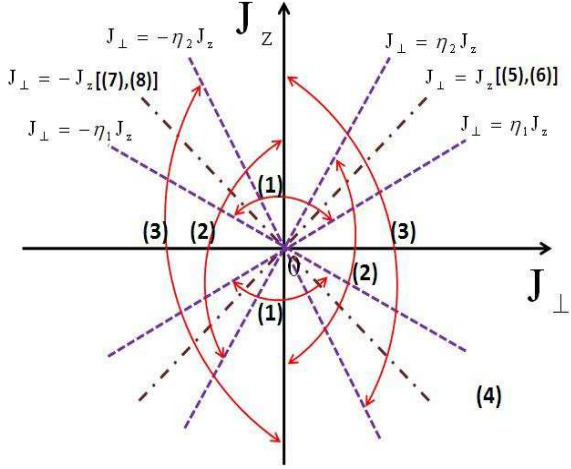}
\caption{(Color online) The stable parameter regimes of the fixed points on $J_\perp-J_z$ plane.  The fixed points
are (1) $\mathbf{n}_a =-\mathbf{n}_b=(0,0,1)$ or $\mathbf{n}_a
=-\mathbf{n}_b=(0,0,-1)$, (2) $\mathbf{n}_a
=-\mathbf{n}_b=(\cos\varphi,\sin\varphi,0)$, (3) $\mathbf{n}_a
=\mathbf{n}_b=(\cos\varphi,\sin\varphi,0)$, (4) $\mathbf{n}_a
=\mathbf{n}_b = (0,0,1)$ or $\mathbf{n}_a =\mathbf{n}_b = (0,0,-1)$,
(5) $\mathbf{n}_a=\mathbf{n}_{b}$, (6)
$\mathbf{n}_a=-\mathbf{n}_{b}$, (7)
$\mathbf{n}_a=(\sin\theta\cos\varphi,
\sin\theta\sin\varphi,\cos\theta)$, while
$\mathbf{n}_b=(\sin\theta\cos\varphi,
\sin\theta\sin\varphi,-\cos\theta)$, (8)
$\mathbf{n}_a=(\sin\theta\cos\varphi,
\sin\theta\sin\varphi,\cos\theta)$, while
$\mathbf{n}_b=(-\sin\theta\cos\varphi,
-\sin\theta\sin\varphi,\cos\theta)$. The fixed point (4) is always stable, which the stable regimes of the other fixed points are indicated by using the curves with arrows at both ends.  }\label{region}
\end{center}
\end{figure}

There exist eight fixed points in our system. Their stability is analyzed in the Appendix. The stable parameter regimes of these fixed points are shown in
Fig.\ref{region}. We specify the fixed point in terms of direction $\mathbf{n}_\alpha$ of the spin vector $\mathbf{S}_\alpha  =S_\alpha\mathbf{n}_\alpha$, that is, $\mathbf{n}_\alpha\equiv
(\sin\theta_\alpha\cos\varphi_\alpha,
\sin\theta_\alpha\sin\varphi_\alpha,\cos\theta_\alpha)$, with  $0\leq\theta_\alpha\leq\pi,0\leq\varphi_\alpha< 2\pi, \alpha=a,b$.

As shown in FIG.~\ref{region}, the fixed points and their stable
regimes are the following.

(1) $\mathbf{n}_a =-\mathbf{n}_b=(0,0,\pm 1)$. One spin is
the parallel to the $z$ direction while the other is antiparallel to
the $z$ direction. This fixed point is stable when
$\eta_1|J_z|>|J_\perp|$, where
$\eta_1\equiv\frac{1}{2}\left(\sqrt{\frac{S_{a}}{S_{b}}}
+\sqrt{\frac{S_{b}}{S_{a}}}\right)$.

(2) $\mathbf{n}_a =-\mathbf{n}_b=(\cos\varphi,\sin\varphi,0)$, where
$0\leq\varphi< 2\pi$. The two spins are antiparallel and are both on the $x-y$ plane. This fixed point is stable if $J_\perp>0$ and $J_\perp>\eta_2J_z$, or $J_\perp<0$ and $J_\perp<\eta_2J_z$, where
$\eta_2\equiv\frac{2S_aS_b}{S_a^2+S_b^2}$.

(3) $\mathbf{n}_a =\mathbf{n}_b=(\cos\varphi,\sin\varphi,0).$ The two spins are parallel and are on the $x-y$ plane. This fixed point
is stable when $J_\perp>0$ and $J_\perp>-\eta_2J_z$, or $J_\perp<0$ and $J_\perp<-\eta_2J_z$.

(4) $\mathbf{n}_a =\mathbf{n}_b = (0,0,\pm 1)$. The two spins are both parallel
or antiparallel to the $z$ direction. This fixed point is always stable.

(5) $\mathbf{n}_a=-\mathbf{n}_{b}$. The two spins are always
antiparallel. This fixed point only exists at $J_\perp=J_z$ and is
stable.

(6)  $\mathbf{n}_a=\mathbf{n}_{b}$. The two spins are
always parallel. This fixed point only exists at $J_\perp=J_z$ and
is stable.

(7)  $\mathbf{n}_a=(\sin\theta\cos\varphi,
\sin\theta\sin\varphi,\cos\theta)$ while
$\mathbf{n}_b=(\sin\theta\cos\varphi,
\sin\theta\sin\varphi,-\cos\theta)$. The $z$ components of
the two spins are opposite. This fixed point only exists at
$J_\perp=-J_z$ and is stable.

(8) $\mathbf{n}_a=(\sin\theta\cos\varphi,
\sin\theta\sin\varphi,\cos\theta)$ while
$\mathbf{n}_b=(-\sin\theta\cos\varphi,
-\sin\theta\sin\varphi,\cos\theta)$. The $xy$ components of the two spins are opposite.  This fixed point only
exists at $J_\perp=-J_z$ and is stable.

It can be seen that
$J_\perp=\pm\eta_1 J_z$ and $J_\perp=\pm\eta_2 J_z$ are bifurcation
points.

\section{Classical ground states \label{classical}}

\begin{figure}
\begin{center}
\includegraphics[height=2.0in,width=3.2in]{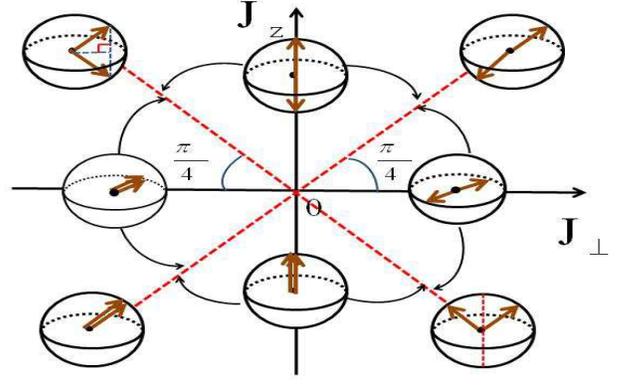}
\caption{(Color online) Classical ground states in different regimes of
parameters $J_\perp$ and $J_z$. They are just the eight fixed
points, but there are differences in the parameter regimes although there are overlaps. (i) $J_z>|J_\perp|$, the fixed point (1)
; (ii) $J_\perp>|J_z|$, the fixed point (2); (iii) $J_\perp<-|J_z|$,
the fixed point (3); (iv) $J_z<-|J_\perp|$, the fixed point (4);
(v)$J_\perp=J_z>0$, the fixed point (5); (vi) $J_\perp=J_z<0$, the
fixed point (6); (vii) $J_\perp=-J_z>0$, the fixed point (7); (viii)
$J_\perp=-J_z<0$, the fixed point (8). } \label{phase}
\end{center}
\end{figure}

As depicted in  Fig.~\ref{phase}, it can be found that  classically  the energy is minimal at fixed point
(1) when $J_z>|J_\perp|$; at fixed point (2) when $J_\perp>|J_z|$; at
fixed point (3) when $J_\perp<-|J_z|$; at fixed point (4) when
$J_z<-|J_\perp|$; at fixed point (5) when $J_\perp=J_z>0$; at fixed point (6) when $J_\perp=J_z<0$; at fixed point (7) when $J_\perp=-J_z>0$; at fixed
point (8) when  $J_\perp=-J_z<0$.

If $S_a=S_b$, we have $\eta_1=\eta_2=1$; therefore the parameter regimes of the
bifurcation points  are completely the same as those of the  classical ground states, respectively. But if $S_a \neq S_b$, there are differences although there are overlap regimes.

\section{Classical evolution \label{evolution} }

Because the Hamiltonian conserves
$S_{az}+S_{bz}$, we study the dynamical evolution of
$S_{az}-S_{bz}$ and $(\varphi_a-\varphi_b)/2$ for some given values
of $S_{az}+S_{bz}$. When $S_{az}=S_a$ while $S_{bz}=S_b$, or
$S_{az}=-S_a$ while $S_{bz}=-S_b$, the system is at the fixed point
(1).  When $S_{az}=0$, $S_{bz}=0$ while
$(\varphi_a-\varphi_b)/2=\pi/2$, the system is at the fixed point
(2). When $S_{az}=0$, $S_{bz}=0$ while $(\varphi_a-\varphi_b)/2=0$ or
$\pi$, the system is at the fixed point (3).

\begin{figure}
\begin{center}
\includegraphics[height=2in,width=3.2in]{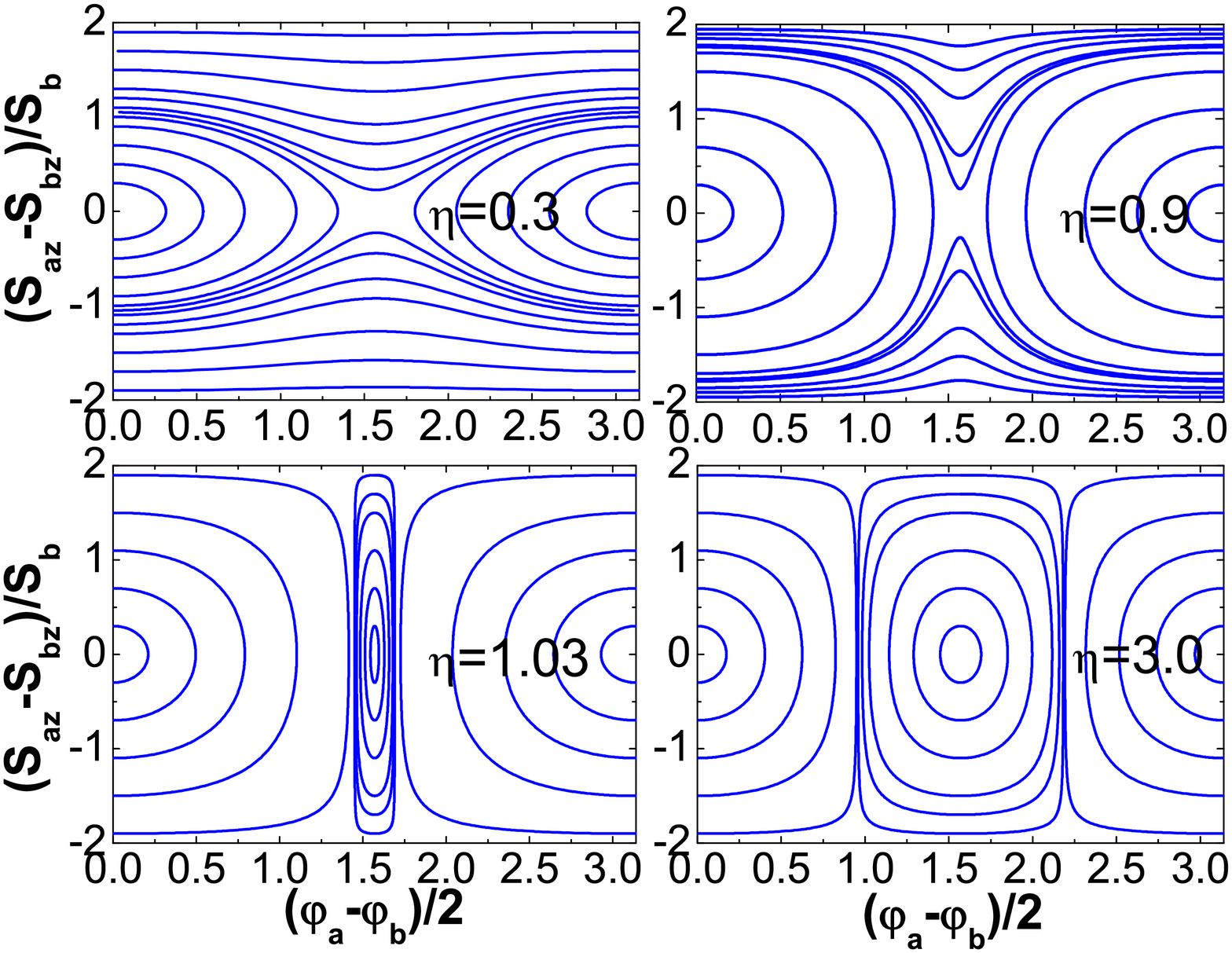}
\caption{(Color online) The evolution trajectories on the plane of  $(S_{az}-S_{bz})/S_{b}$
and $(\varphi_a-\varphi_b)/2$, for various  values of
$S_a=S_b$. Here $\eta \equiv J_\perp/J_z$,
$S_{az}+S_{bz}=0$.}\label{classical1}
\end{center}
\end{figure}

We have studied the evolution trajectories near
fixed points (1), (2) and (3) under various values of $S_a=S_b$.
As shown in Fig.~\ref{classical1}, when
$J_\perp/J_z<1$, fixed points (1) and (3) are stable while fixed
point (2) is unstable; when $J_\perp/J_z>1$, fixed points (2) and
(3) are stable while fixed point (1) is unstable. This conclusion is reached by considering that a fixed point is stable if
the evolution trajectories are  loops around a fixed point, otherwise it
is unstable.

\begin{figure}
\begin{center}
\includegraphics[height=2in,width=3.2in]{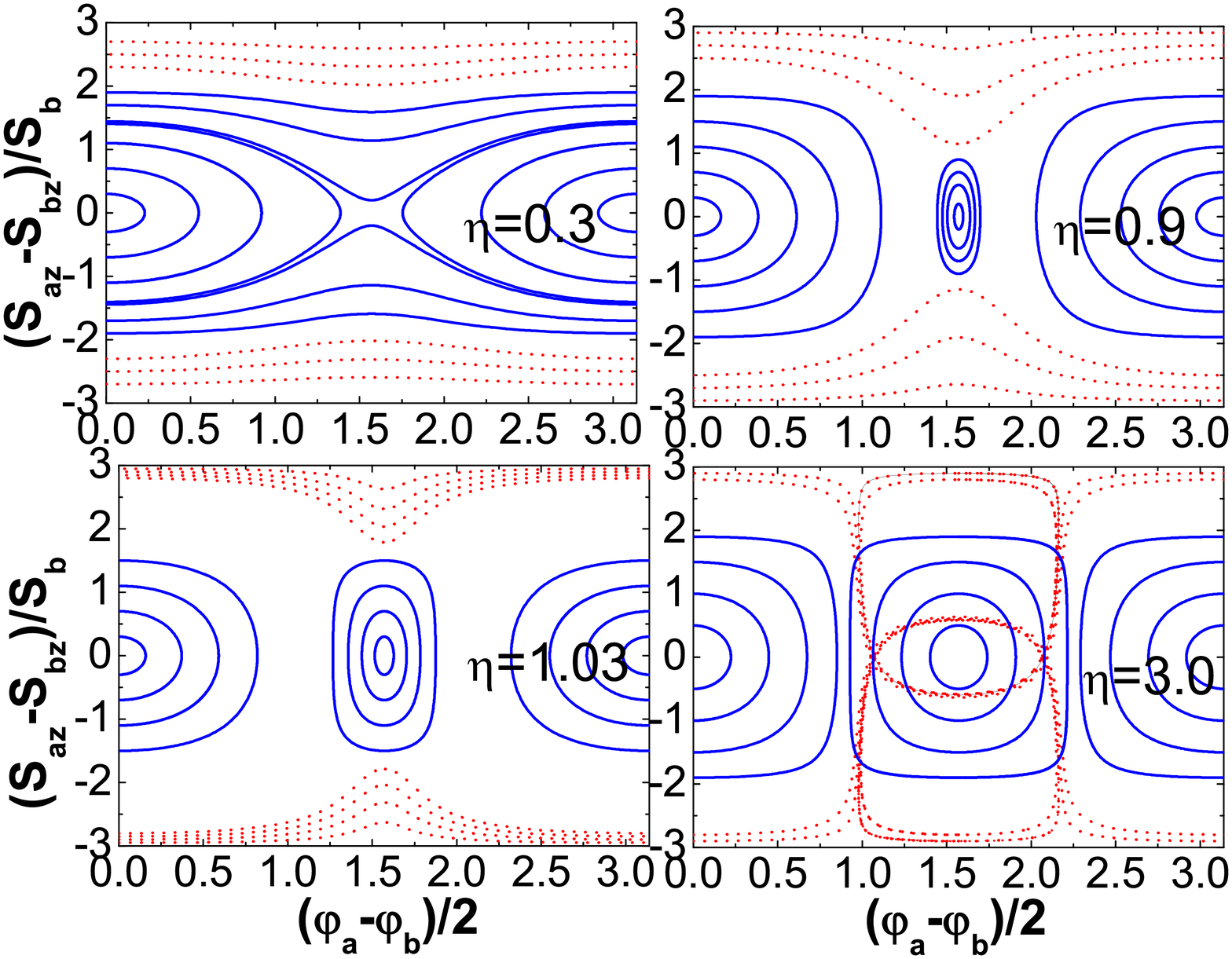}
\caption{(Color online) The evolution trajectories on the plane of  $(S_{az}-S_{bz})/S_{b}$
and $(\varphi_a-\varphi_b)/2$, for various values of
$S_a=2S_b$. Here $\eta \equiv J_\perp/J_z$. The solid lines describe the
dynamics for $S_{az}+S_{bz}=0$ and the dot lines describe the
dynamics for $S_{az}+S_{bz}=\pm(S_a-S_b)$.}\label{classical2}
\end{center}
\end{figure}

We have also
studied the case of $S_a\neq S_b$. As shown in Fig.~\ref{classical2}
for $S_a=2S_b$, the evolution trajectories are different from the
case of $S_a=S_b$. For $J_\perp/J_z=0.9$ and for $J_\perp/J_z=1.03$,
three fixed points are all stable.

Note that all the results of the numerical
simulation are consistent with the above theoretical analysis.

\section{Quantum evolution with disentangled initial state \label{quantum} }

To simulate a quantum process closest to classical evolution, we choose as
the initial state a disentangled state, which can always be written as
\begin{equation}
\begin{split}
|\psi\rangle=(e^{-i\varphi_{a}/2}{\rm
cos}\frac{\theta_{a}}{2}|\uparrow\rangle_{a} +e^{i\varphi_{a}/2}{\rm
sin}\frac{\theta_{a}}{2}|\downarrow\rangle_{a})^{N_{a}}\\
\otimes (e^{-i\varphi_{b}/2}{\rm
cos}\frac{\theta_{b}}{2}|\uparrow\rangle_{b} +e^{i\varphi_{b}/2}{\rm
sin}\frac{\theta_{b}}{2}|\downarrow\rangle_{b})^{N_{b}}\\
=|S_a\mathbf{n}_a\rangle \otimes |S_b\mathbf{n}_b\rangle.
\label{classical state}
\end{split}
\end{equation}
where $|\uparrow\rangle_{\alpha}$ denotes the spin state of a single atom of species $\alpha$, while $|S_\alpha\mathbf{n}_\alpha\rangle$ represents the state of the total spin of species $\alpha$.  In this state, the spin components of each species are similar to classical spins; that is,  $\langle \hat{S}_{\alpha x}\rangle=S_\alpha{\rm
sin}\theta_{\alpha}{\rm cos}\varphi_{\alpha}$, $\langle
\hat{S}_{\alpha y}\rangle=S_\alpha{\rm sin}\theta_{\alpha}{\rm
sin}\varphi_{\alpha}$, and $\langle \hat{S}_{\alpha
z}\rangle=S_\alpha{\rm cos}\theta_{\alpha}$ $(\alpha=a,b)$. Moreover, we choose  $\theta_{\alpha}$ and $\varphi_{\alpha}$ in such a way that $\langle \hat{\mathbf{S}}_{\alpha }\rangle$ corresponds to a fixed point in classical dynamics. For
fixed point (1),  $\mathbf{n}_a=-\mathbf{n}_b=(0,0,1)$; thus the initial state is
$|\psi_{(1)}\rangle=|\uparrow\rangle_a^{\otimes N_a}
|\downarrow\rangle_b^{\otimes N_b}$. For fixed point (2),
$\mathbf{n}_a=-\mathbf{n}_b=(\cos\varphi,\sin\varphi,0)$; thus the initial state is
$|\psi_{(2)}\rangle=(\frac{1}{\sqrt{2}}{\rm
e}^{-i\varphi/2}|\uparrow\rangle_a+\frac{1}{\sqrt{2}}{\rm
e}^{i\varphi/2}|\downarrow\rangle_a)^{\otimes
N_a}(\frac{-i}{\sqrt{2}}{\rm
e}^{-i\varphi/2}|\uparrow\rangle_b+\frac{i}{\sqrt{2}}{\rm
e}^{i\varphi/2}|\downarrow\rangle_b)^{\otimes N_b}$. This is so because
$\hat{S}_{\alpha z}=\hat{N}_{\alpha\uparrow}-\frac{N_\alpha}{2}
=\frac{N_\alpha}{2}-\hat{N}_{\alpha\downarrow}$, where
$\hat{N}_{\alpha\sigma}$ is the number of  atoms of species $\alpha$ with spin $\sigma$ ($\alpha= a, b$; $\sigma = \uparrow, \downarrow$).

For each initial state $|\psi\rangle$ in the form of (\ref{classical state}), we evaluate \begin{equation}
\langle \hat{S}_{\alpha z}(t)\rangle= \langle \psi|e^{i\hat{{\cal H}}t} \hat{S}_{\alpha z}e^{-i\hat{{\cal H}}t}|\psi \rangle,
\end{equation}
whose  evolution actually represents the change of the distribution of
the atoms of species $\alpha$  in the two pseudospin states.

\begin{figure}
\begin{center}
\includegraphics[height=2in,width=3.2in]{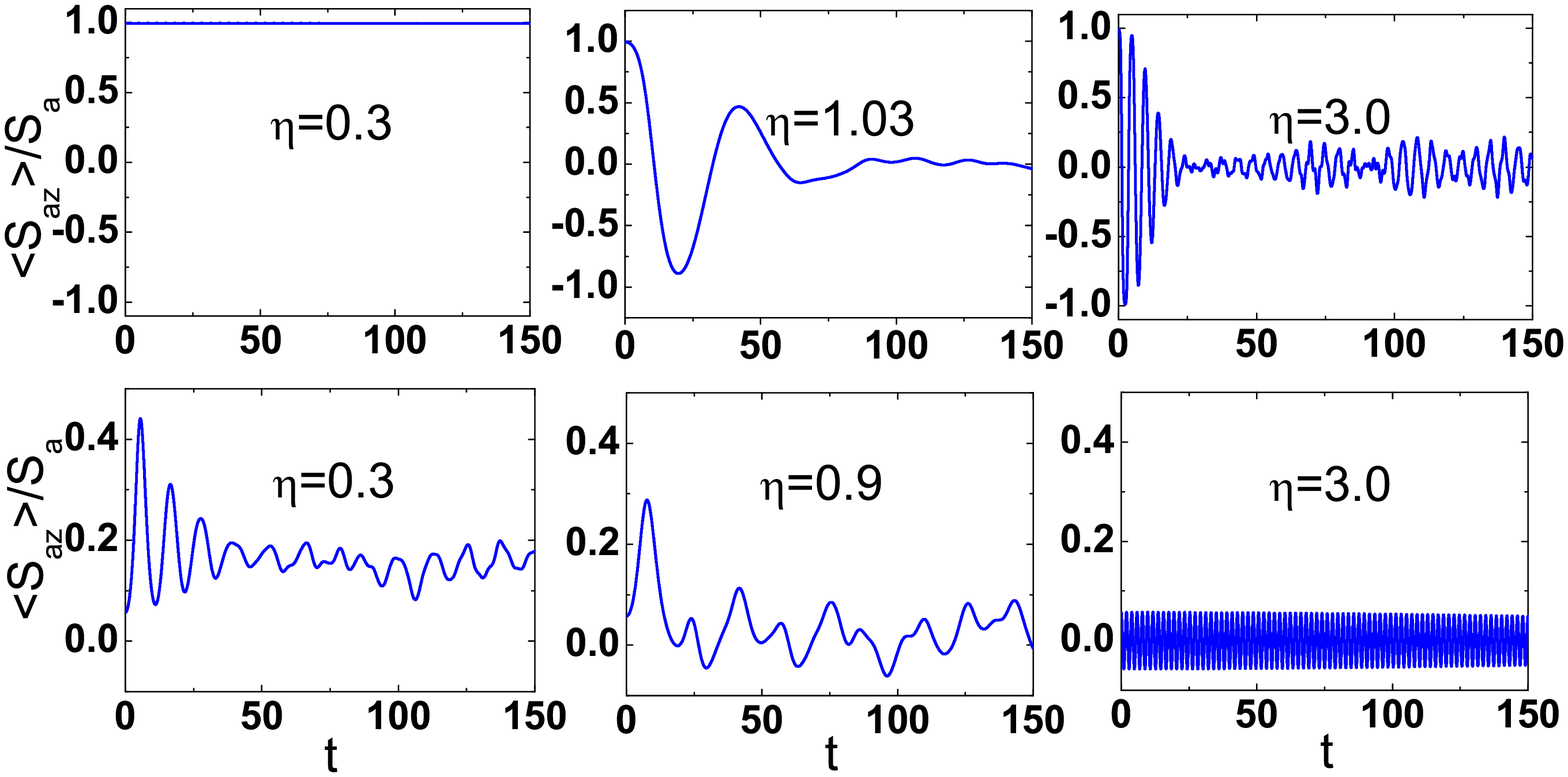}
\caption{(Color online) Quantum dynamics of $\langle
S_{az}\rangle/S_a$ for $S_a=S_b=300$ and various values of $\eta \equiv J_\perp/J_z$. The
figures on the upper line exhibit the stability of fixed point
(1). The figures on the second line exhibit the stability of
fixed point (2). The unit of $t$ is $1/J_z$.}\label{quantum1}
\end{center}
\end{figure}

\begin{figure}
\begin{center}
\includegraphics[height=2in,width=3.2in]{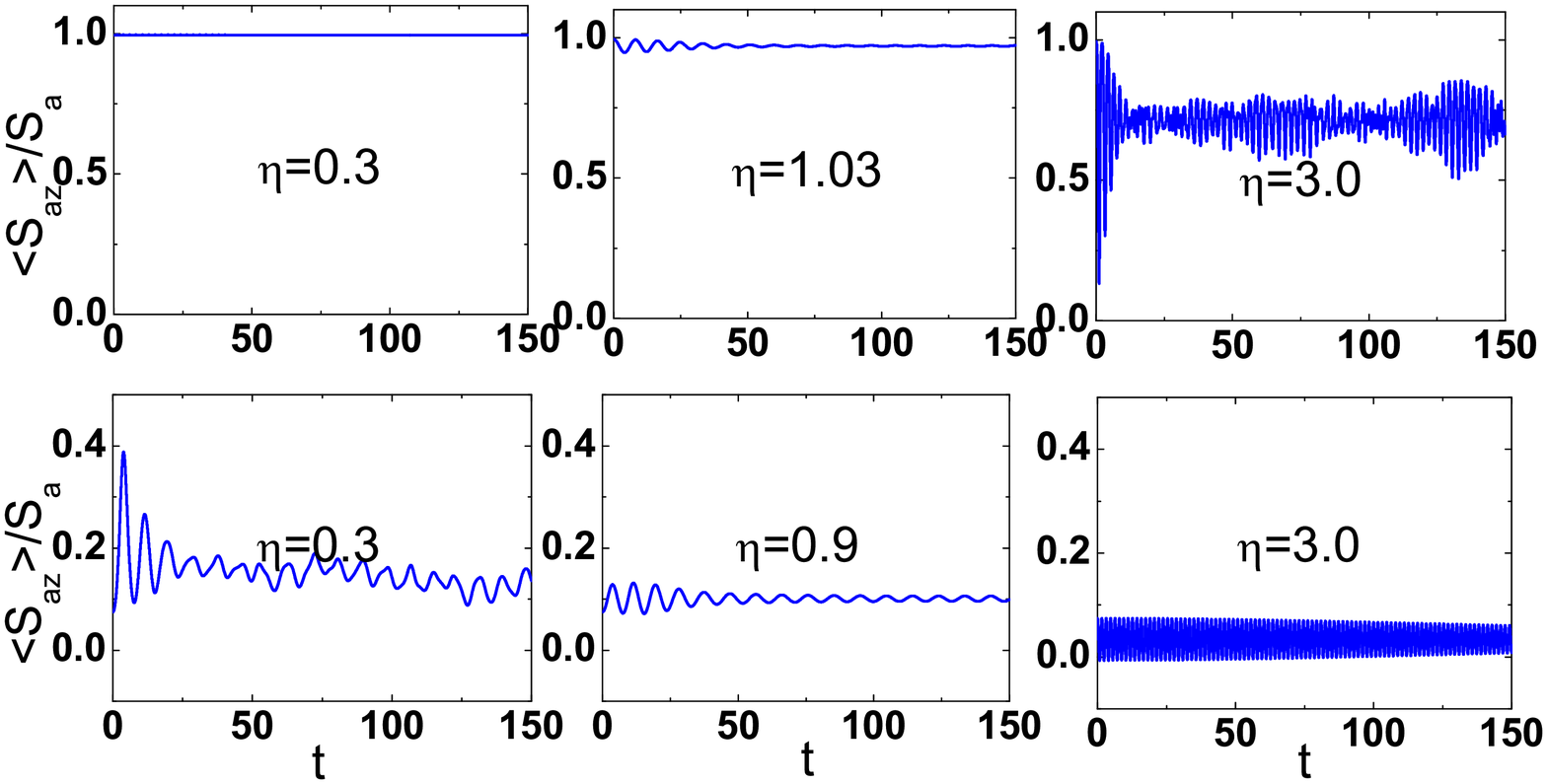}
\caption{(Color online) Quantum dynamics of $\langle
S_{az}\rangle/S_a$ for $S_a=2S_b=300$ and various values of $\eta \equiv J_\perp/J_z$.  The
figures on the upper line exhibit the stability of fixed point
(1). The figures on the second line exhibit the stability of
fixed point (2). The unit of $t$ is $1/J_z$.}\label{quantum2}
\end{center}
\end{figure}

We choose the same initial conditions as in the classical case in last  section  to  start the
quantum dynamics. It  is found that the
classification of the stability of the classical dynamics  still applies. The result is shown in Fig.~\ref{quantum1}  for the case of $S_a=S_b$,  and in Fig.~\ref{quantum2} for the case of $S_a=2S_b$.

The reason why quantum dynamics under the disentangled initial state is so close to classical one is the following. In Heisenberg picture, the quantum equations of motion (\ref{motion1}) reduce to the classical ones (\ref{motionclassical}),
with $\langle S_{\alpha i}\rangle$, ($i=x,y,z$), substituting the corresponding classical spin variable.

\section{Bifurcation and Entanglement \label{entanglement}}

The ground state can always be written as $|G_{S_z}\rangle=\sum
f(m,S_z)|S_a,m\rangle_a|S_b,S_z-m\rangle_b$, where the interspecies entanglement can be quantified as $-\sum_{m}f^2(m,S_z){\rm
log}_{2S_b+1}f^2(m,S_z)$ \cite{shi}. It has been shown that when
$S_a=S_b=S$, the entanglement of the ground state is maximal at
$J_\perp=J_z$, where the ground state is
$(\sqrt{2S+1})^{-1}\sum_{m=-S}^{S}(-1)^m
|S,m\rangle_a|S,-m\rangle_b$   \cite{shi}. Using the transformation $U=e^{i\pi S_{bz}}$,  we can obtain the ground state at $J_\perp=-J_z$ as
$(\sqrt{2S+1})^{-1}\sum_{m=-S}^{S}|S_a,m\rangle_a|S_b,-m\rangle_b$, which is also maximally entangled.

When $S_a=S_b$, there is only one bifurcation point at $J_\perp=J_z$ between the
fixed points (1) and (2); there is also only one bifurcation point at $J_\perp=-J_z$  between the fixed points (1) and  (3). Each of these
bifurcation points corresponds to a maximally entangled quantum
ground state.

However, when $S_a\neq S_b$, $\eta_1\neq\eta_2$, there are
two bifurcation points $J_\perp=\eta_1J_z$ and $J_\perp=\eta_2J_z$ between the
fixed points (1) and (2).  Similarly,   there are
two bifurcation points $J_\perp=-\eta_1 J_z$ and $J_\perp=-\eta_2J_z$ between the
fixed points (1) and (3).

Numerical calculations indicate that in consistency with the classical ground states, the total $z$-component spin exhibits the following features.  When $J_z>0$ while $\eta<1$, $S_z=\pm (S_a-S_b)$.   When $J_z>0$ while  $\eta>1$, $S_z=p$, with $p=0$ if $S_a-S_b$ is an integer while $p=\pm 1/2$ if $S_a-S_b$ is a half integer.

\begin{figure}
\begin{center}
\includegraphics[height=2.0in,width=3.2in]{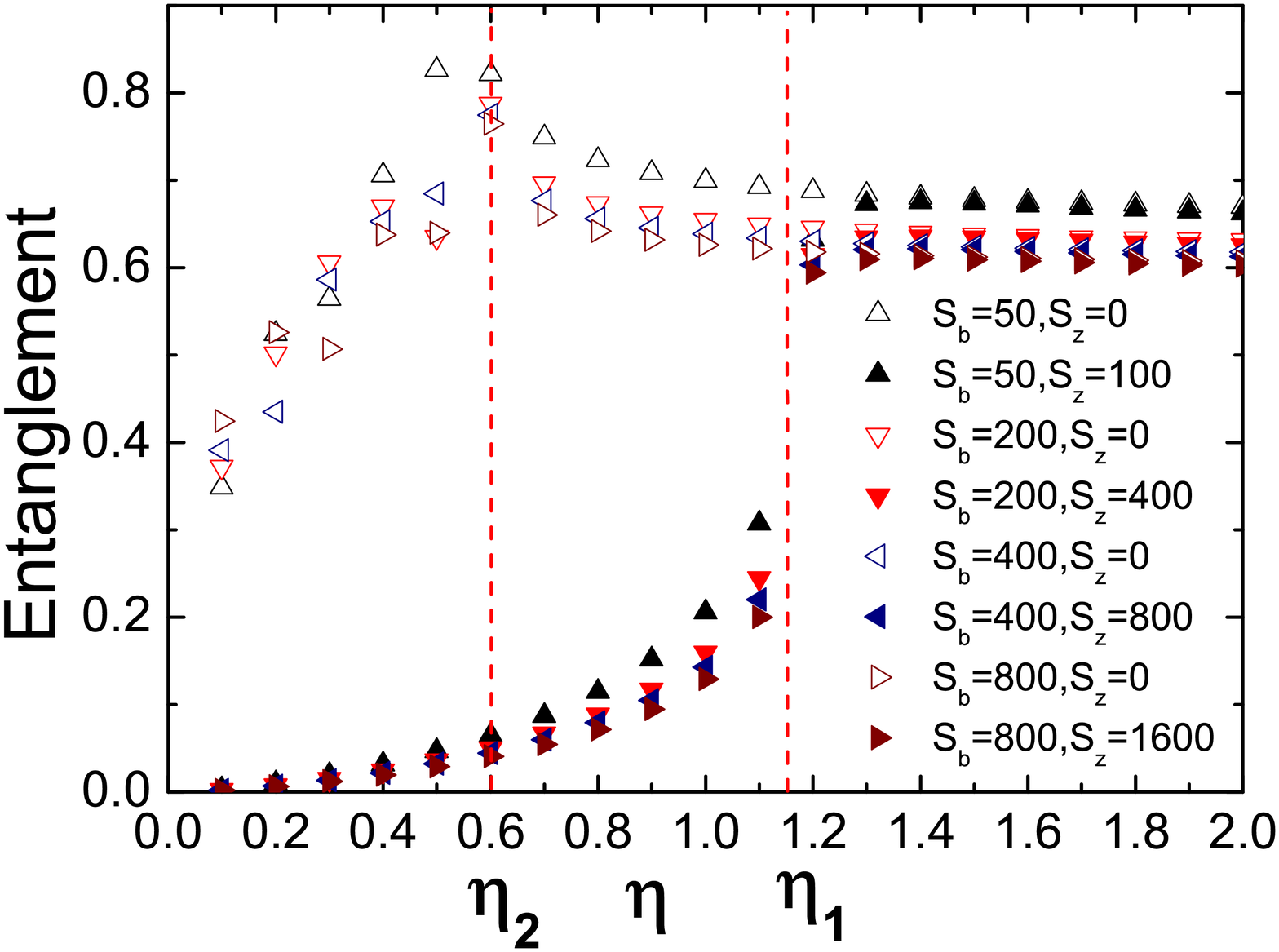}
\caption{(Color online) The numerical result of the entanglement entropy  of the quantum states of $S_{z}=0$ and $S_{z}=S_a-S_b$, as a function of $\eta$ . Here $S_a=3S_b$, $J_z >0$.}\label{groundstate}
\end{center}
\end{figure}

Numerical results of the entanglement entropy of the states with $S_z=S_a-S_b$ and $S_z=0$, varying with $\eta$, are shown in  Fig.\ref{groundstate} for some integer values of $S_b$ and $S_a=3S_b$.
For  $\eta<1$, the ground state is the one with $S_z=S_a-S_b$, whose entanglement values are plotted as empty triangles. For $\eta>1$, the ground state is the one  with $S_z=0$, whose entanglement values are plotted as filled triangles. Therefore, there is a discontinuity of entanglement in passing $\eta =1$, where both states are degenerate ground states.

According to the bifurcation analysis discussed above,  when $S_a=3S_b$, the two bifurcation points are  $\eta_1 =\frac{1}{2}(\sqrt{1/3}+\sqrt{3})\approx 1.1547$ and $\eta_2= 0.6$. As indicated in FIG.{\ref{groundstate}}, the entanglement entropy of the states with $S_{z}=0$ and $S_z=S_a-S_b$ is maximal at $\eta_1$ and $\eta_2$, respectively, and  decreases rapidly in deviating from each of them, with the decrease more rapid for $\eta$ larger than the maximal point.

Therefore, when $S_a\neq S_b$, the quantum state corresponding to each  classical fixed point still possesses maximal entanglement at the parameter point where the fixed point bifurcates. However, this quantum state is not the quantum ground state. In other words, the entanglement of the quantum
ground state at each bifurcation point is not maximal anymore.

\section{Analytical solutions of the quantum ground states  \label{ground} }

\begin{figure}
\begin{center}
\includegraphics[height=2.0in,width=3.0in]{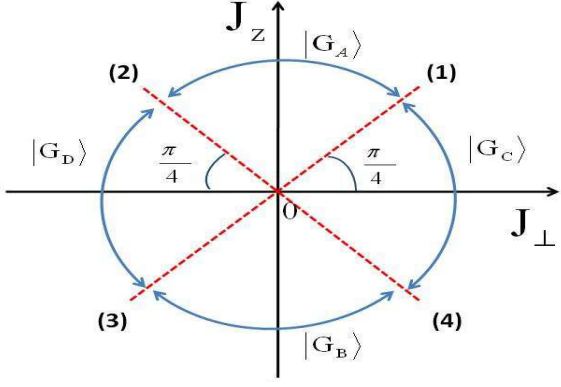}
\caption{(Color online) The quantum ground states in all the parameter regimes.   $|G_i\rangle$ $(i=A,B,C,D)$ represent the ground states in the four bulk regimes, and are given in Secs. \ref{a} to \ref{d}. (1) to (4) represent the four boundaries, in which the ground states are described in Sec.~\ref{boundaries}.}\label{qg}
\end{center}
\end{figure}

We now proceed to analytically find out the quantum ground states on all $J_\perp-J_z$ parameter regimes, by using effective Hamiltonians which describe deviations from the classical ground state in each parameter regime. All the ground states are summarized in Fig.~\ref{qg}. Regimes A ($J_z> |J_\perp|$) and B ($J_z<-|J_\perp|$) both correspond to $|\xi_s-\xi_d| > |\xi_e| $; i.e., the interspecies spin exchange scattering is quite weak, with A and B differing in whether the equal-spin forward scattering length is larger or smaller than the unequal-spin forward scattering length. Regimes C ($J_\perp>|J_z|$) and D ($J_\perp<-|J_z|$) both correspond to $ |\xi_e| > |\xi_s-\xi_d| $; i.e., the interspecies spin exchange scattering is quite strong, with C and D differing in whether the spin-exchange scattering length is positive or negative.

Regime A ($J_z>J_\perp$) corresponds to $\xi_s-\xi_d > |\xi_e| $; i.e., the interspecies spin exchange scattering is quite weak.

\subsection{$J_z>|J_\perp|$ \label{a}}

As shown in Fig.~2, in this parameter regime, the classical ground state is fixed point (1), i.e., $|S_a,S_a\rangle|S_b,-S_b\rangle$ or $|S_a,-S_a\rangle|S_b,S_b\rangle$.

First we consider the quantum ground state near $|S_a,S_a\rangle|S_b,-S_b\rangle$. One can make
the Holstein-Primarkoff transformation \cite{hol},
\begin{equation}
\begin{split}
&\hat{S}_{a-}=\hat{f}_a^\dagger\sqrt{2S_a-\hat{f}_a^\dagger\hat{f}_a},\quad \hat{S}_{a+}=\sqrt{2S_a-\hat{f}_a^\dagger\hat{f}_a} \hat{f}_a,\\
&\hat{S}_{az}=S_a-\hat{f}_a^\dagger\hat{f}_a,\\ &\hat{S}_{b-}=\sqrt{ 2S_b-\hat{f}_b^\dagger\hat{f}_b}\hat{f}_b,\quad
\hat{S}_{b+}=\hat{f}_b^\dagger\sqrt{ 2S_b-\hat{f}_b^\dagger\hat{f}_b },\\
&\hat{S}_{bz}=\hat{f}_b^\dagger \hat{f}_b-S_b,
\end{split}
\end{equation}
with
$\hat{S}_{\alpha\pm} \equiv \hat{S}_{a x}\pm i \hat{S}_{\alpha y}$,  $\alpha=a,b$,  $\hat{f}_\alpha$ and $\hat{f}_\alpha^\dagger$ being bosonic operators satisfying
$\hat{f}_\alpha|n_\alpha\rangle
=\sqrt{n_\alpha}|n_\alpha-1\rangle$,
$\hat{f}_\alpha^\dagger|n_\alpha\rangle
=\sqrt{n_\alpha+1}|n_\alpha+1\rangle,$
$[\hat{f}_\alpha,\hat{f}_\beta]=0,$
$[\hat{f}_\alpha,\hat{f}_\beta^\dagger]=\delta_{\alpha\beta}$,
where
\begin{eqnarray}
|n_a\rangle & \equiv &
|S_a,S_a-n_a\rangle_a, \\
|n_b\rangle  & \equiv &
|S_b,-S_b+n_b\rangle_b,
\end{eqnarray}
$n_{\alpha}=0,1,2,\cdots, 2S_\alpha$. When $S_\alpha$ is
very large, $\langle \hat{f}_\alpha^+\hat{f}_\alpha\rangle\ll
2S_\alpha$, $\hat{S}_{\alpha -} \approx(2S_a)^{1/2}\hat{f}_{\alpha}^\dagger$,
$\hat{S}_{a z} \hat{S}_{bz}\approx
S_b\hat{f}_a^+\hat{f}_a+S_a\hat{f}_b^+\hat{f}_b-S_aS_b$. Then  the Hamiltonian (\ref{spinhamiltonian}) can be approximated as
\begin{equation}
\begin{split}
\hat{{\cal H}}\approx&-J_zS_aS_b+J_z(S_b\hat{f}_a^\dagger\hat{f}_a+S_a\hat{f}_b^\dagger \hat{f}_b)\\
&+J_\perp\sqrt{S_aS_b}(\hat{f}_a^\dagger\hat{f}_b^\dagger+\hat{f}_b\hat{f}_a), \label{app1}
\end{split}
\end{equation}
Then we make the Bogoliubov transformation
\begin{equation}
\begin{split}
&\hat{f}_c=\sqrt{\frac{\Delta_1+1}{2}}
\hat{f}_a +sgn(J_\perp) \sqrt{\frac{\Delta_1-1}{2}}
\hat{f}_b^\dagger,\\
&\hat{f}_d= sgn(J_\perp) \sqrt{\frac{\Delta_1-1}{2}}\hat{f}_a^\dagger+\sqrt{\frac{\Delta_1+1}{2}}
\hat{f}_b,
\end{split} \label{cd}
\end{equation}
where $sgn(J_\perp)$ is the sign of $J_\perp$,
$\Delta_1 \equiv \frac{J_z(S_a+S_b)}{\sqrt{J_z^2(S_a+S_b)^2-4J_\perp^2S_aS_b}}.$
When $S_a=S_b$ and $J_z=J_\perp$, $\Delta_1 \pm 1$ should be $1$.
Hamiltonian(\ref{app1}) becomes
\begin{equation}
\hat{{\cal H}}_A= \epsilon_{1c} \hat{f}_c^\dagger\hat{f}_c + \epsilon_{1d} \hat{f}_d^\dagger\hat{f}_d +E_{10},\label{app2}
\end{equation}
where $E_{10} \equiv
-J_zS_aS_b+\frac{J_z(\Delta_1-1)}{2}(S_a+S_b)-|J_\perp|\sqrt{(\Delta_1^2-1)
S_aS_b}$, $\epsilon_{1c} \equiv
\frac{J_z(\Delta_1-1)}{2}S_a+\frac{J_z(\Delta_1+1)}{2}S_b-
|J_\perp|\sqrt{(\Delta_1^2-1)S_aS_b}$, $\epsilon_{1d} \equiv
\frac{J_z(\Delta_1+1)}{2}S_a+\frac{J_z(\Delta_1-1)}{2}S_b
-|J_\perp|\sqrt{(\Delta_1^2-1)S_aS_b}$.
Thus the energy spectrum is
$E_A(n_c,n_d)= \epsilon_{1c} n_c + \epsilon_{1d} n_d +E_{10}$,
where $n_c$ and $n_d$ are nonnegative integer numbers. For
$J_z>|J_\perp|$, $\epsilon_{1c}$ and $\epsilon_{1d}$ are always
positive; therefore the ground-state energy is
$E_1(0,0)= E_{10}$.
When $J_z \rightarrow |J_\perp|$,
$E_0(0,0)$ approaches   $-S_b(S_a+1)$, which is the  the exact ground-state energy at $J_z = |J_\perp|$.

Like the original Hamilton (\ref{spinhamiltonian}), the effective  Hamiltonian (\ref{app1}) also conserves the $z$ component of the total spin. Therefore any of its eigenstates can be written as
\begin{equation}
|\psi_1(n_c,n_d)\rangle=\sum_m
g_1(n_c,n_d,m)|S_a,m\rangle_a|S_b,S_z-m\rangle_b,
\end{equation}
where ${\rm max}(-S_a,S_z-S_b)\leq m\leq {\rm min}(S_a,S_z+S_b)$, $g_1(n_c,n_d,m)$ is the expansion coefficient, and $S_z$
is the total $z$ component of the spin system. Using (\ref{cd}) and considering that
$|\psi_1(n_c,n_d)\rangle$ is an eigenstate of both
$\hat{f}_c^\dagger\hat{f}_c$ and $\hat{f}_d^\dagger\hat{f}_d$, with eigenvalues $n_c$ and $n_d$ respectively,  we obtain
\begin{equation}
n_c-n_d  =  S_a-S_b-S_z.
\end{equation}
For the ground state $|\psi_1(0,0)\rangle$, $S_z=S_a-S_b$.

It is easy to find the ground state $|\psi_1(0,0)\rangle$ of (\ref{app1}) from
$\hat{f}_c|\psi_1(0,0)\rangle=0$,
\begin{widetext}
\begin{eqnarray}
|\psi_1(0,0)\rangle& =& D\sum_{m=S_a-2S_b}^{S_a} \left[-sgn(J_\perp) \sqrt{\frac{\Delta_1+1}{\Delta_1-1}}\right]^m
|S_a,m\rangle_a|S_b, S_a-S_b-m\rangle_b,\\
&=& D\sum_{m=-S_b}^{S_b} \left[-sgn(J_\perp) \sqrt{\frac{\Delta_1+1}{\Delta_1-1}}\right]^{S_a-S_b-m} |S_a, S_a-S_b-m\rangle_a|S_b, m \rangle_b,
\label{ground1}
\end{eqnarray}
\end{widetext}
where $
D\equiv [(\frac{\Delta_1+1}{\Delta_1-1})^{S_a-S_b}
\frac{(\Delta_1+1)^{2S_b+1}-(\Delta_1-1)^{2S_b+1}}{2(\Delta_1^2-1)^{S_b}}]^{-1/2}$
is the normalization coefficient.
When $J_\perp\rightarrow 0$,  $|G_A\rangle\rightarrow |S_a,S_a\rangle|S_b,-S_b\rangle$,  which
is an  exact ground state of the Hamiltonian (\ref{spinhamiltonian})
with $J_\perp=0$ and $J_z>0$.

The excited states of (\ref{app1}) can be obtained  by the action of  $\hat{f}_c^\dagger$ and $\hat{f}_d^\dagger$ on the ground state $|\psi_1(0,0)\rangle$. With $S_a > S_b$, $\epsilon_{1c} < \epsilon_{1d} $, for a given  $S_{z}$, the lowest excited state is  $|\psi_1(n_c,0)\rangle$ if $S_{z} < S_a-S_b$ and is $|\psi_1(0,n_d)\rangle$ if $S_{z} > S_a-S_b$.  These two excited states can be written as
\begin{widetext}
\begin{equation}
 |\psi_1(n_c,0)\rangle=D_c\exp(-i\zeta\pi \hat{S}_{az})\sum_{n=0}^{n_c}\sum_{m=S_a-2S_b-n}^{S_a-n_c}
(-1)^{m+n}(\sqrt{\frac{\Delta_1+1}{\Delta_1-1}})^{m+2n}
C_{n_c}^{n}\sqrt{C_{S_a-m}^{n_c}}|S_a,m\rangle_a|S_b,S_a-S_b-n_c-m\rangle_b, \label{14}
\end{equation}
\begin{equation}
|\psi_1(0,n_d)\rangle=D_d\exp(-i\zeta\pi \hat{S}_{az})\sum_{m=S_a-2S_b+n_d}^{S_a}(-\sqrt{\frac{\Delta_1+1}{\Delta_1-1}})^{m}
\sqrt{C_{S_a+n_d-m}^{n_d}}|S_a,m\rangle_a|S_b,S_a-S_b+n_d-m\rangle_b, \label{15}
\end{equation}
\end{widetext}
where $\zeta=0$ if $J_\perp >0$ while $\zeta=1$ if $J_\perp <0$;  $D_c$ and $D_d$ are the normalization constants, and $C_{n}^{m}$ is the binomial coefficient.

Now we consider the ground state close to the other classical degenerate ground state $|S_a,-S_a\rangle|S_b,S_b\rangle$, in a way similar to the above. The Holstein-Primarkoff transformation is
\begin{equation}
\begin{split}
&\hat{S}_{b-}'=\hat{f}_b'^\dagger\sqrt{2S_b-\hat{f}_b'^\dagger\hat{f}_b'},\quad \hat{S}_{b+}'=\sqrt{2S_b-\hat{f}_b'^\dagger\hat{f}_b'} \hat{f}_b',\\
&\hat{S}_{bz}'=S_b-\hat{f}_b'^\dagger\hat{f}_b',\\ &\hat{S}_{a-}'=\sqrt{ 2S_a-\hat{f}_a'^\dagger\hat{f}_a'}\hat{f}_a',\quad
\hat{S}_{a+}'=\hat{f}_a'^\dagger\sqrt{ 2S_a-\hat{f}_a'^\dagger\hat{f}_a' },\\&
\hat{S}_{az}'=\hat{f}_a'^\dagger \hat{f}_a'-S_a,
\end{split}
\end{equation}
where the bosonic operators $f_a'$ and $f_b'$ act  on
\begin{eqnarray}
|n_a'\rangle & \equiv &
|S_a,-S_a+n_a'\rangle_a, \\
|n_b'\rangle  & \equiv &
|S_b,S_b-n_b'\rangle_b.
\end{eqnarray}
Thus one obtains a Hamiltonian
\begin{equation}
\begin{split}
\hat{{\cal H}_A'}&\approx-J_zS_aS_b+J_z(S_b\hat{f}_a'^\dagger\hat{f}_a'+S_a\hat{f}_b'^\dagger \hat{f}_b)' \nonumber \\
& +J_\perp\sqrt{S_aS_b}(\hat{f}_a'^\dagger\hat{f}_b'^\dagger+\hat{f}_b'\hat{f}_a'), \\ &
=\epsilon_{1c} \hat{f}_c'^\dagger\hat{f}_c' + \epsilon_{1d} \hat{f}_d'^\dagger\hat{f}_d' +E_{10},
\end{split}\label{app123}
\end{equation}
where
\begin{equation}
\begin{split}
&\hat{f}_c'=\sqrt{\frac{\Delta_1+1}{2}}
\hat{f}_a' +sgn(J_\perp) \sqrt{\frac{\Delta_1-1}{2}}
\hat{f}_b'^\dagger,\\
&\hat{f}_d'= sgn(J_\perp) \sqrt{\frac{\Delta_1-1}{2}}\hat{f}_a'^\dagger+\sqrt{\frac{\Delta_1+1}{2}}
\hat{f}_b'.
\end{split} \label{cd2}
\end{equation}
Therefore the eigenstates can be written as
\begin{equation}
|\psi_1'(n_c',n_d')\rangle=\sum_m
g_1(n_c',n_d',m)|S_a,m\rangle_a|S_b,S_z-m\rangle_b,
\end{equation}
with the constraint
\begin{equation}
n_c'-n_d'=S_a-S_b+S_z.
\end{equation}
For the ground state $|\psi_1'(0,0)\rangle$, $S_z=S_b-S_a$.
\begin{widetext}
\begin{eqnarray}
|\psi_1'(0,0)\rangle
 =D'\sum_{m=-S_b}^{S_b} \left[-sgn(J_\perp) \sqrt{\frac{\Delta_1+1}{\Delta_1-1}}\right]^m |S_a, S_b-S_a-m\rangle_a|S_b,m\rangle_b,
\label{ground12}
\end{eqnarray}
\end{widetext}
where $
D'\equiv[\frac{(\Delta_1+1)^{2S_b+1}-(\Delta_1-1)^{2S_b+1}}{2(\Delta_1^2-1)^{S_b}}]^{-1/2}.$
When $J_\perp\rightarrow 0$,  $|G_A\rangle\rightarrow |S_a,-S_a\rangle|S_b,S_b\rangle$,  which
is an exact ground state of the Hamiltonian (\ref{spinhamiltonian})
with $J_\perp=0$ and $J_z>0$.

The excited states of (\ref{app123}) can be obtained  by the action of  $\hat{f}_c'^\dagger$ and $\hat{f}_d'^\dagger$ on the ground state $|\psi_1'(0,0)\rangle$. With $S_a > S_b$, $\epsilon_{1c} < \epsilon_{1d} $, for a given  $S_{z}$, the lowest excited state is  $|\psi_1'(n_c',0)\rangle$ if $S_{z} > S_b-S_a$ and is $|\psi_1'(0,n_d')\rangle$ if $S_{z} < S_b-S_a$. The explicit expressions of  $|\psi_1'(n_c,0)\rangle$ and $|\psi_1'(0,n_d)\rangle$ are like (\ref{14}) and (\ref{15}) for
$|\psi_1(n_c,0)\rangle$ and $|\psi_1(0,n_d)\rangle$, with $|S_a,m\rangle_a$, $|S_b,S_a-S_b-n_c-m\rangle_b$ and $|S_b,S_a-S_b+n_d-m\rangle$ replaced as $|S_a,-m\rangle_a$, $|S_b,-S_a+S_b+n_c+m\rangle_b$ and $|S_b,-S_a+S_b-n_d+m\rangle$, respectively.

It is important to note that $|\psi_1(0,0)\rangle$  and $|\psi_1(0,0)\rangle$ are orthogonal unless $S_a=S_b$.
Therefore, when $S_a\neq S_b$,  the ground states are doubly degenerate ones $|\psi_1(0,0)\rangle$ and $ |\psi'_1(0,0)\rangle$ at each parameter point in this regime.

When  $S_a=S_b$, $\gamma \equiv \langle\psi_1(0,0)|\psi'_1(0,0)\rangle =\frac{2(2S_b+1)(\Delta_1^2-1)^{S_b}}{(\Delta_1+1)^{2S_b+1}-(\Delta_1-1)^{2S_b+1}}
$;  hence we must find the ground state in their two-dimensional subspace. Clearly $\langle\psi_1(0,0)|\hat{\cal H}|\psi_1(0,0)\rangle = \langle\psi'_1(0,0)|\hat{\cal H}|\psi'_1(0,0)\rangle  \approx E_{10}$. $\langle\psi_1(0,0)|\hat{\cal H}|\psi'_1(0,0)\rangle  =\langle\psi_1'(0,0)|\hat{\cal H}|\psi_1(0,0)\rangle \approx E_{10}\gamma$.  Consequently, the ground state is found to be
\begin{equation}
|G_A(S_a=S_b)\rangle=\frac{1}{\sqrt{2}}[|\psi_1(0,0)\rangle +|\psi_1'(0,0)\rangle],
\end{equation}
with energy $E_{10}(1+\gamma)$.
The energy of $\frac{1}{\sqrt{2}}[|\psi_1(0,0)\rangle-|\psi_1'(0,0)\rangle]$ is $E_{10}(1-\gamma)$.

When $J_\perp$ and $J_z$ approach the boundary $J_z=-J_\perp>0$ from the regime of  $|G_A\rangle$,   $|G_A\rangle$ approaches  $e^{i\pi S_{az}}|S_a-S_b,\pm(S_a-S_b)\rangle$.
When  $J_\perp$ and $J_z$ approach the boundary $J_z=J_\perp>0$ from the regime of   $|G_A\rangle$,  $|G_A\rangle$ approaches $|S_a-S_b,\pm (S_a-S_b)\rangle$.

\subsection{$J_z<-|J_\perp|$ }

In this parameter regime,
the classical ground states are
$|S_a,S_a\rangle|S_b,S_b\rangle$,  in which the two spins are both along the $z$ direction, and
$|S_a,-S_a\rangle|S_b,-S_b\rangle$,  in which the two spins are both along the $-z$ direction.

Consider the ground state close to
$|S_a,S_a\rangle_a|S_b,S_b\rangle_b$. We make the Holstein-Primarkoff
transformation
$\hat{S}_{\alpha -} =\hat{h}_\alpha^\dagger\sqrt{2S_\alpha-\hat{h}_\alpha^\dagger \hat{h}_\alpha}$,  $\hat{S}_{\alpha +}=\sqrt{2S_\alpha-\hat{h}_\alpha^\dagger\hat{h}_\alpha}\hat{h}_\alpha,$
$\hat{S}_{\alpha z}=S_\alpha-\hat{h}_\alpha^\dagger\hat{h}_\alpha$, where $\hat{h}_\alpha$ and $\hat{h}_\alpha^\dagger$ are  bosonic operators, now with $|n_\alpha\rangle \equiv
|S_\alpha,S_\alpha-n_\alpha\rangle_\alpha$, where $n_\alpha=0,1\cdots, 2S_\alpha$. When $S_\alpha$ is
very large, $\langle \hat{h}_\alpha^+\hat{h}_\alpha\rangle\ll
2S_\alpha$, $\hat{S}_{\alpha -}\approx(2S_\alpha)^{1/2}\hat{h}_\alpha^\dagger$, $\hat{S}_{az}\hat{S}_{bz}\approx
S_aS_b-S_b\hat{h}_a^+\hat{h}_a-S_a\hat{h}_b^\dagger\hat{h}_b$.
Then
the Hamiltonian ({\ref{spinhamiltonian}}) becomes
\begin{equation}
\begin{split}
\hat{{\cal H}}_B=&J_zS_aS_b-J_z(S_b
\hat{h}_a^\dagger\hat{h}_a+S_a\hat{h}_b^\dagger\hat{h}_b)\\
&+J_\perp\sqrt{S_aS_b}(\hat{h}_a^\dagger\hat{h}_b+\hat{h}_b^\dagger\hat{h}_a).
\label{appb}
\end{split}
\end{equation}

We define another two bosonic operators,
\begin{equation}
\begin{split}
&\hat{h}_c=-sgn(J_\perp)\sqrt{\frac{1-\Delta_2}{2}}\hat{h}_a+\sqrt{\frac{1+\Delta_2}{2}}\hat{h}_b,\\
&\hat{h}_d=sgn(J_\perp)\sqrt{\frac{1+\Delta_2}{2}}\hat{h}_a+\sqrt{\frac{1-\Delta_2}{2}}\hat{h}_b,
\end{split}\label{b2}
\end{equation}
where
$
\Delta_2 \equiv \frac{J_z(S_a-S_b)}{\sqrt{J_z^2(S_a-S_b)^2+4J_\perp^2S_aS_b}}.$ Then the Hamiltonian (\ref{spinhamiltonian}) becomes
\begin{equation}
\hat{{\cal H}}_B= \epsilon_{2c} \hat{h}_c^\dagger\hat{h}_c + \epsilon_{2d} \hat{h}_d^\dagger\hat{h}_d +E_{20},\label{b3}
\end{equation}
where
$E_{20} \equiv
J_zS_aS_b$, $\epsilon_{2c} \equiv -[\frac{J_z(1+\Delta_2)}{2}S_a+
\frac{J_z(1-\Delta_2)}{2}S_b+|J_\perp|\sqrt{(1-\Delta_2^2)S_aS_b}]$, $\epsilon_{2d} \equiv-[\frac{J_z(1-\Delta_2)}{2}S_a+\frac{J_z(1+\Delta_2)}{2}S_b-|J_\perp|
\sqrt{(1-\Delta_2^2)S_aS_b}].$
The energy spectrum is $
E_B(n_c,n_d)=\epsilon_{2c} n_c + \epsilon_{2d} n_d +E_{20}$.
Thus the  ground state is $|S_a,S_a\rangle|S_b,S_b\rangle$.

The excited state $|\psi_2(n_c,n_d)\rangle$ of (\ref{appb}) can be obtained by the action of $\hat{h}_c^{\dagger}$ and $\hat{h}_d^\dagger$ on the ground state $|S_a,S_a\rangle|S_b,S_b\rangle$. It is obvious that $\epsilon_{2c}$ is always larger than $\epsilon_{2d}$, for a given $S_z$, the lowest excited state is $|\psi_2(n_c,0)\rangle$, with
\begin{equation}
n_c=S_a+S_b-S_z,
\end{equation}
\begin{widetext}
\begin{equation}
|\psi_2(n_c,0)\rangle=D_{2c}\exp(-i\zeta\pi S_{az})\sum_{m=max(n_c-2S_b,0)}^{n_c}(-\sqrt{\frac{1-\Delta_2}{1+\Delta_2}})^{m}\sqrt{C_{n_c}^m}|S_a,S_a-m\rangle_a|S_b,S_b-n_c+m\rangle_b,\label{exb1}
\end{equation}
\end{widetext}
where $\zeta=0$ if $J_\perp >0$ while $\zeta=1$ if $J_\perp <0$,
and $D_{2c}=[\frac{(1+\Delta_2)^{n_c}(1-\Delta_2)^{n_0}-(1+\Delta_2)^{n_0-1}(1-\Delta_2)^{n_c+1}}{2\Delta_2(\Delta_2+1)^{n_c+n_0-1}}]^{-1/2}$, where $n_0=max(n_c-2S_b,0)$.

Using the same method, we  obtain the approximate Hamiltonian $\hat{{\cal H}}_B'$
close to the other classical ground state $|S_a,-S_a\rangle|S_b,-S_b\rangle$, which turns out to be the quantum ground state. The Holstein-Primarkoff
transformation is now
$\hat{S}_{\alpha +} =\hat{h}_\alpha'^\dagger\sqrt{2S_\alpha-\hat{h}_\alpha'^\dagger \hat{h}_\alpha'}$,
$\hat{S}_{\alpha -}=\sqrt{2S_\alpha-\hat{h}_\alpha'^\dagger{\hat{h}_\alpha'}}{\hat{h}_\alpha'},$
$\hat{S}_{\alpha z}=\hat{h}_\alpha'^\dagger{\hat{h}_\alpha'}-S_\alpha$, where ${\hat{h}_\alpha'}$ and $\hat{h}_\alpha'^\dagger$ are  bosonic operators, with $|n'_\alpha\rangle \equiv
|S_\alpha,n'_\alpha-S_\alpha\rangle_\alpha$. Thus we have Eqs. (\ref{appb}) to (\ref{b3}) with primed operators.
In this set of eigenstates,
the lowest excited states for a given $S_z$ are $|\psi_2'(n_c',0)\rangle$, with
\begin{equation}
n_c'=S_z+S_a+S_b.
\end{equation}
$|\psi_2'(n_c,0)\rangle$ are like (\ref{exb1}), with $|S_a,S_a-m\rangle_a$ and $|S_b,S_b-n_c+m\rangle_b$ replaced by $|S_a,-S_a+m\rangle_a$ and $|S_b,-S_b+n_c-m\rangle_b$.

$|\psi_2(0,0)\rangle$  and $|\psi_2'(0,0)\rangle$ are orthogonal, hence are just the doubly  degenerate ground states in this regime,  and can be written as
\begin{equation}
|G_B\rangle = |S_a+S_b, \pm (S_a+S_b)\rangle.
\end{equation}

When  $J_\perp$ and $J_z$ approach the boundary $J_z=J_\perp<0$
from the regime of $|G_B\rangle$,  $|G_B\rangle$ approaches $|S_a+S_b, \pm(S_a+S_b)\rangle$.
When  $J_\perp$ and $J_z$ approach the boundary $J_\perp=-J_z> 0$ from the regime of $|G_B\rangle$, $|G_B\rangle$ approaches $|S_a+S_b, \pm(S_a+S_b)\rangle$.

\subsection{$J_\perp>|J_z|$}

In this parameter regime,
it is convenient to rewrite the Hamiltonian as
\begin{equation}
\begin{split}
{\cal H}&=J_\perp \sqrt{(S_a^2-S_{az}^2)(S_b^2-S_{bz}^2)}\cos(\varphi_a-\varphi_b)+J_z S_{az}S_{bz},
\end{split}
\end{equation}
where   $\varphi_{\alpha}$ ($\alpha=a,b$) is the azimuthal angle.

In the vicinity of the classical ground state, $S_{az}\sim 0$,
$S_{bz}\sim 0$ , $\varphi_a-\varphi_b \sim \pi$, for simplicity we
define $\varphi_a-\varphi_b=\varphi_{ab}+\pi$. Therefore
\begin{equation}
\begin{split}
{\cal H}_C &\approx -J_\perp \sqrt{S_aS_b(S_a+1)(S_b+1)}+\frac{J_\perp^2-J_z^2}{4(J_\perp\xi_{+}-J_z)}S_{z}^2\\
&+\frac{J_\perp\xi_{+}-J_z}{4}(S_{2z}-\frac{J_\perp\xi_{-}}{J_\perp\xi_{+}-J_z}S_{1z})^2
\\&+
2J_\perp\sqrt{S_aS_b(S_a+1)(S_b+1)}(\frac{\varphi_{ab}}{2})^2 \label{approxh1},
\end{split}
\end{equation}
where  $S_{2z} \equiv S_{az}-S_{bz}$,
$\xi_{+} \equiv \frac{1}{2}(\frac{S_a}{S_b}+\frac{S_b}{S_a})$,
$\xi_{-} \equiv \frac{1}{2}(\frac{S_a}{S_b}-\frac{S_b}{S_a})$.

$S_{z}$ commutes with ${\cal H}$, and is thus a constant of motion. Then
$\hat{P}_2\equiv \hat{S}_{2z}-\frac{J_\perp\xi_{-}}{J_\perp\xi_{+}-J_z}\hat{S}_{z}$ and  $\hat{X}_{2}\equiv \hat{\varphi}_{ab}/2$ are conjugate variables, as $\hat{S}_{\alpha z}$ and
$\hat{\varphi}_{\alpha}$ are  canonically conjugate variables.
The   Hamiltonian is then similar to that of a harmonic oscillator.
The energy spectrum is thus
\begin{equation}
\begin{split}
&E_3(n,S_z)=-J_\perp\sqrt{S_aS_b(S_a+1)(S_b+1)}\\
&
+\frac{J_\perp^2-J_z^2}{4(J_\perp\xi_{+}-J_z)}S_z^2\\
&+(n+\frac{1}{2})\sqrt{2J_\perp(J_\perp\xi_{+}-J_z)\sqrt{S_a(S_a+1)S_b(S_b+1)}}\\
&\approx -J_\perp\sqrt{S_aS_b(S_a+1)(S_b+1)}+\frac{J_\perp^2-J_z^2}{4(J_\perp\xi_{+}-J_z)}S_z^2\\
&+(n+\frac{1}{2})\sqrt{2J_\perp(J_\perp\xi_{+}-J_z)S_a S_b}\label{energy2},\\
\end{split}
\end{equation}
where $n$ is the quantum number of the harmonic oscillator.
The eigenstate  for $n=0$ can be written as
\begin{equation}
|\psi_3(0,S_z)\rangle= Z(S_z) \sum_m
f(m,S_z)|S_a,m\rangle_a|S_b,S_z-m\rangle_b,
\label{3},
\end{equation}
where
$f(m,S_z)=(-1)^m \exp[-\sqrt{\frac{J_\perp\xi_{+}-J_z}{2J_\perp
S_aS_b}}m(m-\frac{J_\perp\frac{S_a}{S_b}-J_z}{J_\perp\xi_{+}-J_z}S_z)],$
with $\rm{max}(-S_a,S_z-S_b)\leq m \leq
\rm{min}(S_a,S_z+S_b)$;
$Z(S_z) \equiv \frac{1}{\sqrt{\sum_m |f(m,S_z)|^2}}$ is the normalization coefficient.

The ground state is   thus
\begin{equation}
|G_C\rangle=|\psi_3(0,p)\rangle= Z(p)\sum
f(m,p)|S_a,m\rangle_a|S_b,p-m\rangle_b,
\label{approxstate2}
\end{equation}
where $p=0$ if $S_a-S_b$ is an integer, while $p=\pm 1/2$ if $S_a-S_b$ is a half integer.

For $S_a=S_b=S$ and $J_\perp\gg J_z$, we have calculated the entanglement entropy of
$|G_C\rangle$,
\begin{equation}
\begin{split}
{\cal E}(|G_C\rangle)&=-\sum[Z(p)f(m,p)]^2{\rm log}_{2S+1}[Z(p)f(m,p)]^2.
\end{split}
\end{equation}
It is evaluated that when $S\rightarrow \infty$,
${\cal E}(|G_C\rangle)\approx  1/2,$
which is very large.

When  $J_\perp$ and $J_z$ approach the boundary $J_z=J_\perp >0$ from  the regime of  $|G_C\rangle$, it approaches $|S_a-S_b,p\rangle$, where $p=0$ if $S_a-S_b$ is an integer while $p=\pm 1/2$ if   $S_a-S_b$ is a half integer.
When  $J_\perp$ and $J_z$ approach the boundary $J_z=-J_\perp <0$ from  the regime of  $|G_C\rangle$, it
approaches $e^{i\pi S_{az}}|S_a+S_b, p \rangle$.

\subsection{$J_\perp<-|J_z|$ \label{d}}

The energy spectrum for $J_\perp<0$ can be obtained by using  ${\cal H}(J_\perp,J_z)=UH(-J_\perp,J_z)U^{\dagger}$, where  $U\equiv e^{i\pi
S_{az}}$. Therefore, in the regime  $J_\perp<-|J_z|$, the energy spectrum is  also given by Eq.~(\ref{energy2}).

The ground state is thus
\begin{equation}
\begin{split}
|G_D\rangle
 &= U|G_C\rangle
\\&
= Z(p)\sum |f(m,p)| |S_a,m\rangle|S_b,p-m\rangle,
\end{split}
\end{equation}
with $\rm {max}(-S_a,S_z-S_b)\leq m \leq
\rm{min}(S_a,S_z+S_b)$.

Obviously the entanglement entropy of $|G_D\rangle$ is the same as that of $|G_C\rangle$, with $J_\perp$ reversing its sign.

When  $J_\perp$ and $J_z$ approach the boundary $J_z=-J_\perp>0$ from  the regime of  $|G_D\rangle$, it approaches $e^{i\pi S_{az}}|S_a-S_b,p\rangle$.
When  $J_\perp$ and $J_z$ approach the boundary $J_z=J_\perp<0$ from  the regime of  $|G_D\rangle$, it approaches $|S_a+S_b,p\rangle$.

\subsection{The ground states on the four parameter boundaries \label{boundaries} }

When $J_z=J_\perp>0$, the Hamiltonian
(\ref{spinhamiltonian}) is ${\cal H}=J_z \mathbf
{\hat{S}}_a\cdot\mathbf{\hat{S}}_b$, the degenerate ground states are
$|S_a-S_b,S_z\rangle =\sum_m
g(S_a-S_b,S_z,m)|S_a,m\rangle|S_b,S_z-m\rangle$,  with $S_z=S_b-S_a,  \cdots,
S_a-S_b$. Here $g(S_a-S_b,S_z,m)$ is the Clebsch-Gordan coefficient.

When $J_z=-J_\perp>0$, the degenerate ground states are
$e^{i\pi \hat{S}_{az}}|S_a-S_b,S_z\rangle=\sum_m (-1)^m
g(S_a-S_b,S_z,m)|S_a,m\rangle|S_b,S_z-m\rangle$, with  $S_z=S_b-S_a,  \cdots,
S_a-S_b$.

When $J_z=J_\perp<0$, the
ground states are $|S_a+S_b,S_z\rangle=\sum_m
g(S_a+S_b,S_z,m)|S_a,m\rangle|S_b,S_z-m\rangle$, with $S_z=-S_a-S_b$,
$\cdots$, $S_a+S_b$.

When $J_z=-J_\perp<0$, the ground states are
$e^{i\pi \hat{S}_{az}}|S_a+S_b,S_z\rangle=\sum_m (-1)^m
g(S_a+S_b,S_z,m)|S_a,m\rangle|S_b,S_z-m\rangle$, with $S_z=-S_a-S_b$,
$\cdots$, $S_a+S_b$.

The boundaries are where quantum phase transition take place.
We have known that the ground states $|G_A\rangle$,    $|G_B\rangle$, $|G_C\rangle$, $|G_D\rangle$, in the four regimes discussed in previous subsections, depend on the values of $J_z$ and $J_\perp$. Starting as a ground state in one of these regimes (see FIG.~\ref{qg}), when $J_z$ and $J_\perp$ adiabatically approach each boundary regime, the ground state always approaches one of the degenerate ground states on the boundary. In entering the other regime across the boundary, the ground state restarts from another one of the degenerate ground states on the boundary.

\subsection{Comparison with the numerical results}

As each eigenstate for $J_\perp<0$ can be obtained by acting $e^{i\pi \hat{S}_{az}}$ on an eigenstate for $J_\perp=|J_\perp|$, we only need to consider the  half of the parameter space with $J_\perp \geq 0$.

In this half parameter space,  We have calculated the dependence of the entanglement on $1/\eta\equiv J_z/J_\perp$, using the ground states analytically obtained above in regimes $A$, $C$, and $B$.  We compare these analytical results with the numerical results. The reason of choosing $1/\eta$ rather than $\eta$ is because $J_z=0$ in the middle of the half parameter space. In this half parameter space, regime B is $1/\eta < -1$, regime C is $-1 < 1/\eta < -1$, while regime A is $1/\eta > 1$.

\begin{figure}
\begin{center}
\includegraphics[height=2.0in,width=3.2in]{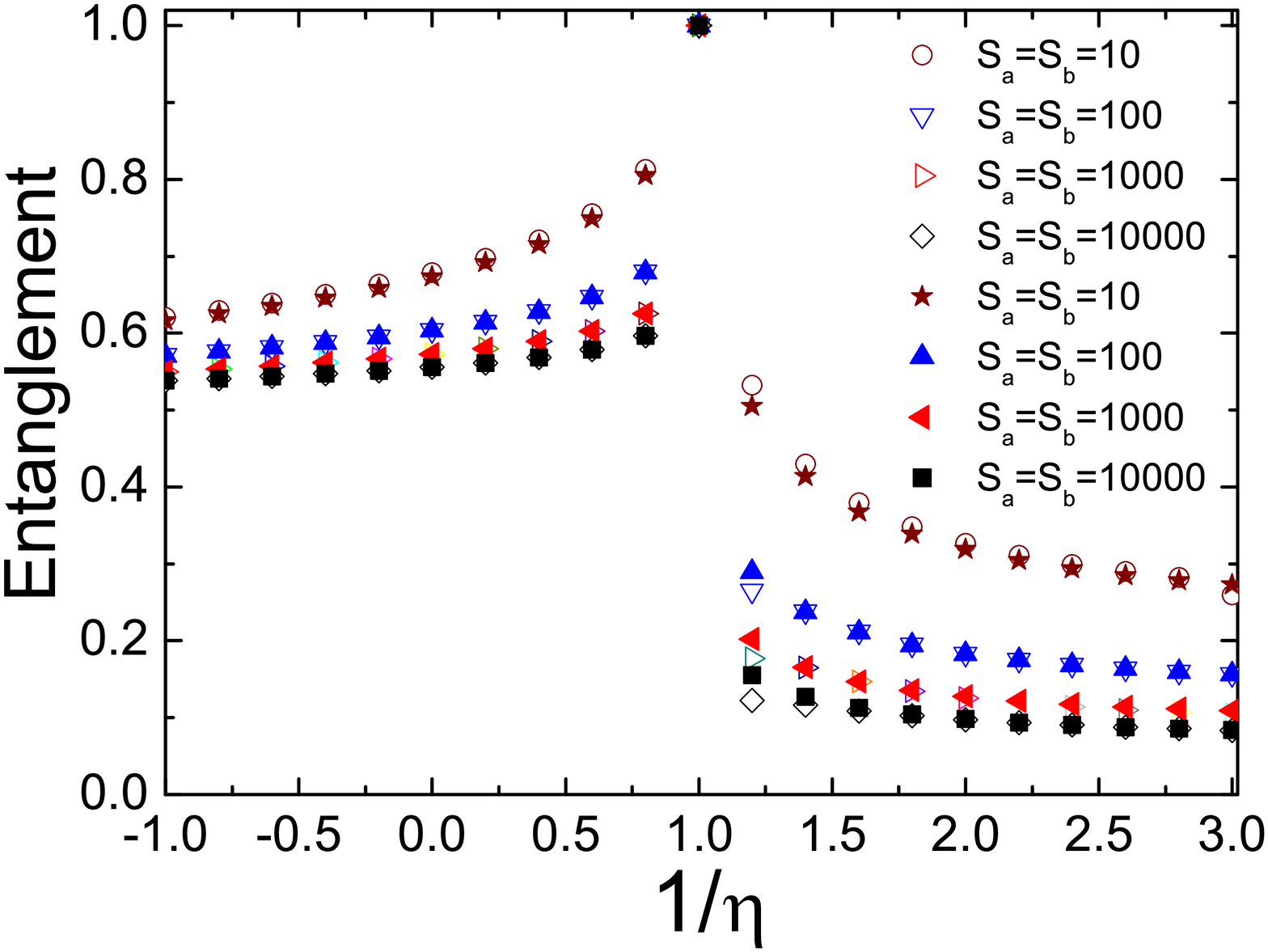}
\caption{(Color online) The entanglement entropy of the ground state as a function of $1/\eta \equiv =J_z/J_\perp$. The filled symbols describe the analytical results.  The empty symbols describe  the numerical results.}\label{compare1}
\end{center}
\end{figure}

Figure \ref{compare1} shows the the entanglement in the ground states for different values of $S_a=S_b$ for $1/\eta > -1$.  Neglected is regime B, i.e.  $1/\eta < -1$, as the ground state is exactly $|S_a,S_a\rangle|S_b,S_b\rangle$ or $|S_a,-S_a\rangle|S_b,-S_b\rangle$, without entanglement. Figure \ref{compare1} clearly indicates excellent fitting between the analytical results in this section and the numerical results.

\begin{figure}
\begin{center}
\includegraphics[height=2.0in,width=3.2in]{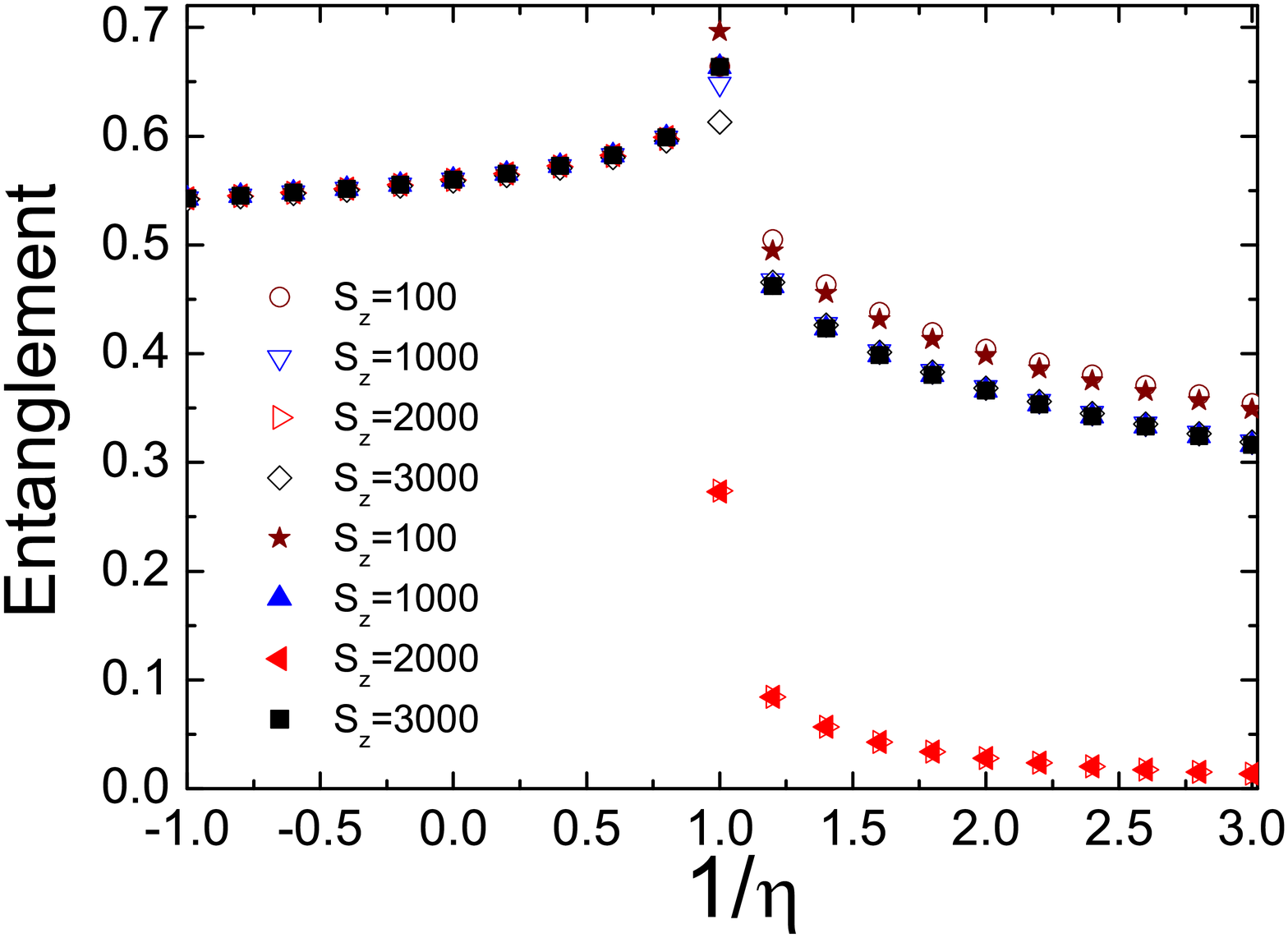}
\caption{(Color online) The entanglement entropy of the lowest energy excited states for different values of $S_z$ under $S_a=12000$ and $S_b=10000$,  as a function $1/\eta=J_z/J_\perp$. Here we show the regime $\eta > -1$, in which the low-energy excited states are with small magnitudes of $S_z$.  The filled symbols describe the analytical results.  The empty symbols describe the numerical results.   }\label{compare2}
\end{center}
\end{figure}

\begin{figure}
\begin{center}
\includegraphics[height=2.0in,width=3.2in]{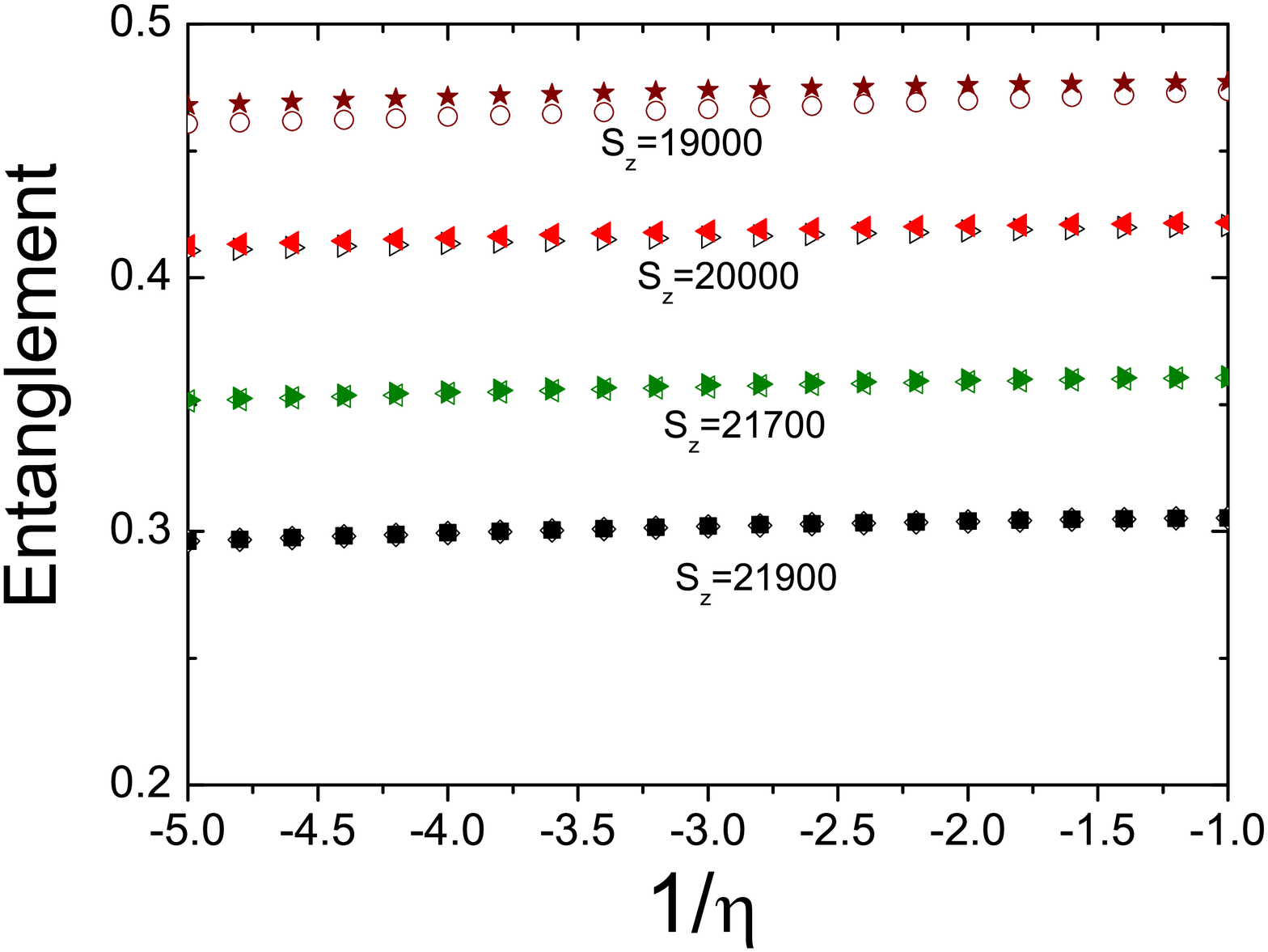}
\caption{(Color online) The entanglement entropy of the quantum states as a function $1/\eta=J_\perp/J_z$, for different values of $S_z$ under $S_a=12000$ and $S_b=10000$ in  regime B ($1/\eta < -1$), where the low energy excited states have large magnitudes of $S_z$. The filled symbols describe the analytical results.  The empty symbols describe the  numerical results.   } \label{figb}
\end{center}
\end{figure}

Excellent fitting between our analytical results and the numerical results are also obtained for excited states. We have calculated the lowest energy states of different values of $S_z$ for $S_a=12000$ and $S_b=10000$. Figure \ref{compare2} shows the
the regime $1/\eta > -1$, i.e., regimes C and A, while  Fig.\ref{figb} shows
the regime $1/\eta < -1$, i.e., regime B. The reason for this separation is that the low-energy excited state in regime B is with large magnitudes of  $S_z$, while those in regimes C and A are with small  magnitudes of  $S_z$.

In conclusion,  our analytical results fit the numerical results very well.

\section{Summary \label{summary}}

In this paper, we considered a binary mixture of two species of pseudo-spin-$\frac{1}{2}$ atoms with interspecies spin exchange in the absence of an external potential, and extended the study of its  ground states to the whole parameter space of the two effective spin coupling strengths. Meanwhile, this provides a model of studying the relation between the classical model and quantum ground states.

We first analyzed the corresponding classical Hamiltonian. We  found the  fixed points of the classical dynamics, and discussed their stability situation both analytically and numerically. The bifurcations were discussed.

The classical evolution can be reproduced in quantum dynamics if starting from an initial state which is disentangled between the two species, as we have demonstrated.

In the case that the atom numbers of the two species are equal, we  confirmed in our system the previous claim that a classical fixed point  bifurcation  corresponds to maximal entanglement in the quantum ground state. Moreover, we find the result that when the two atom numbers are unequal, the entanglement of the quantum
ground state at the parameter point of the bifurcation is not maximal, while the state corresponding to the fixed point that bifurcates indeed possesses maximal entanglement at  that parameter point.

A quantum ground state can be regarded as the classical ground state with quantum fluctuations. This perspective leads to  solutions of the ground states in all parameter regimes, by obtaining an effective Hamiltonian near each classical ground state. Using entanglement entropy as the quantity characterizing the ground states, we find that the analytical results fit the numerical results very well. We have made many detailed discussions.

Our work establishes EBEC as a system manifesting connections between classical dynamics and quantum behavior.

\acknowledgements

We thank Li Ge and Bailin Hao for useful discussion.
This work was supported by the National Science Foundation of China (Grant No. 11074048)  and the Ministry of Science and Technology of China (Grant No. 2009CB929204).

\appendix *

\section{Classical fixed points}

\begin{table*}
\caption{\label{table1} Stable regimes of the fixed
points.
$\eta_1 \equiv \frac{1}{2}\left(\sqrt{\frac{S_{a}}{S_{b}}}+\sqrt{\frac{S_{b}}{S_{a}}}\right)$,
$\eta_2\equiv \frac{2S_{a}S_{b}}{S_{a}^2+S_{b}^{2}}$.}
\begin{tabular}{lll}
\hline \hline No.& \qquad\qquad\qquad Fixed Points &\quad Stable regions\\
 1 & $\mathbf{n}_a = -\mathbf{n}_b=(0,0,\pm 1)$ & \quad $\eta_1|J_z|>|J_\perp|$\\
 2 & $\mathbf{n}_a =-\mathbf{n}_b=
(\cos\varphi,\sin\varphi,0)$ &\quad $J_\perp>0$,$J_\perp>\eta_2J_z$
or $J_\perp<0,
J_\perp<\eta_2J_z$\\
 3 & $\mathbf{n}_a =\mathbf{n}_b=
(\cos\varphi,\sin\varphi,0)$& \quad$J_\perp>0$, $J_\perp>-\eta_2J_z$
or
$J_\perp<0$, $J_\perp<-\eta_2J_z$\\
 4 & $\mathbf{n}_a =\mathbf{n}_b= (0,0,\pm 1)$ & \qquad\quad All \\
 5 & $\mathbf{n}_a=-\mathbf{n}_{b}$ & $\qquad\quad J_\perp=J_z$\\
 6 & $\mathbf{n}_a=\mathbf{n}_{b}$ & $\qquad\quad J_\perp=J_z$\\
 7 & $\mathbf{n}_a=(\sin\theta\cos\varphi,
\sin\theta\sin\varphi,\cos\theta)$,& $\qquad\quad J_\perp=-J_z$ \\
& $\mathbf{n}_b=(\sin\theta\cos\varphi,\sin\theta\sin\varphi,-\cos\theta)$&\\
 8 &
$\mathbf{n}_a=(\sin\theta\cos\varphi,\sin\theta\sin\varphi,\cos\theta)$,&
$\qquad\quad J_\perp=-J_z$ \\
&$\mathbf{n}_b=(-\sin\theta\cos\varphi,
-\sin\theta\sin\varphi,\cos\theta)$&\\
\hline
\end{tabular}
\end{table*}

For our problem, the desired Lyapunov function will always be
found by defining
\begin{equation}
{\cal L}  =  \gamma_1 {\cal H}+\gamma_2 J_z (S_{az}+S_{bz})^{2}
\label{Liapunov}
\end{equation}
where $\gamma_1$ and $\gamma_2$ are suitably chosen coefficients. It
is clear that $\frac{d \cal L}{dt}=0$.

We find the following fixed points specified by the values of
$\mathbf{S}_\alpha =S_\alpha\mathbf{n}_\alpha$, where
$\mathbf{n}_\alpha\equiv
(\sin\theta_\alpha\cos\varphi_\alpha,\sin\theta_\alpha\sin\varphi_\alpha,\cos\theta_\alpha)$
, $0\leq\theta_\alpha\leq\pi,0\leq\varphi_\alpha< 2\pi, \alpha=a,b$.

(1) $\mathbf{n}_a =-\mathbf{n}_b=(0,0,\pm 1)$; that is, one spin is parallel to
the $z$ direction, the other is antiparallel to the $z$ direction. At these
two point the eigenvalues of ${\cal J}$ are $\mu_{1}=\mu_{2}=0$,
$\mu_{3,4}=\pm\sqrt{\frac{\zeta_{3}-\zeta_{4}}{2}}$,
$\mu_{5,6}=\pm\sqrt{\frac{\zeta_{3}+\zeta_{4}}{2}}$, where
$\zeta_{3}\equiv
2J_\perp^{2}S_{a}S_{b}-J_z^{2}(S_{a}^{2}+S_{b}^{2}), $ $\zeta_{4}
=J_z(S_{a}-S_{b})\sqrt{J_z^{2}(S_{a}+S_{b})^{2}-4J_\perp^{2}S_{a}S_{b}}$.
If we define
$\eta_1=\frac{1}{2}\left(\sqrt{\frac{S_a}{S_b}}+\sqrt{\frac{S_b}{S_a}}\right)$,
when $\eta_1|J_z|<|J_\perp|$, there are eigenvalues with positive
real part, and these two fixed points are unstable. Otherwise, the
stabilization cannot be judged by the eigenvalues. One finds the
Lyapunov function ${\cal L}={\cal
H}-\frac{J_z}{4}(S_{az}+S_{bz})^{2}$, which is minimal for $J_z>0$
(or maximal for $J_z<0$) at each of these two fixed points in the
parameter region $\eta_1|J_z|>|J_\perp|$, where each of these two
fixed points is thus stable.

(2) $\mathbf{n}_a =-\mathbf{n}_b=(\cos\varphi,\sin\varphi,0)$, where
$0\leq\varphi< 2\pi$. The two spins are antiparallel and  both are on the $x-y$
plane. At this point, the eigenvalues of ${\cal J}$ are
$\mu_1=\mu_2=\mu_3=\mu_4=0$, $\mu_{5,6}=\pm\sqrt{2J_\perp J_z
S_{a}S_{b}-J_\perp^2(S_{a}^{2}+S_{b}^{2})}$. If we define
$\eta_2=\frac{2S_{a}S_{b}}{S_{b}^{2}+S_{b}^{2}}$, when
$0<J_\perp<\eta_2 J_z$ or $\eta_2 J_z<J_\perp<0$, some eigenvalues
have positive real part, hence this fixed point is unstable. When
$J_\perp>\eta_2J_z\geq 0$ or $J_\perp<\eta_2J_z\leq 0$, one finds
the Lyapunov function ${\cal L}={\cal H}+\gamma_2
J_z(S_{az}+S_{bz})^{2}$, which is minimal at the fixed point as
$\gamma_2 \rightarrow\infty$. When $J_zJ_\perp<0$, one finds that
${\cal L}=-{\cal H}+\gamma_2J_z(S_{az}+S_{bz})^2$ is minimal at the
fixed point as $\gamma_2 \rightarrow\infty$. Thus this fixed point
is stable if $J_\perp>0$,$J_\perp>\eta_2J_z$ or $J_\perp<0,
J_\perp<\eta_2J_z$.

(3) $\mathbf{n}_a =\mathbf{n}_b=(\cos\varphi,\sin\varphi,0)$; that
is, the two spins are parallel and on the $x-y$ plane.  At this point,
the eigenvalues of ${\cal J}$ are $\mu_1=\mu_2=\mu_3=\mu_4=0$,
$\mu_{5,6}=\pm\sqrt{-J_\perp^{2}(S_{a}^{2}+S_{b}^{2})-2J_\perp J_z
S_{a}S_{b}}$. When $-\eta_2J_z<J_\perp<0$ or $0<J_\perp<-\eta_2J_z$,
some eigenvalues have positive real parts, hence the fixed point is
unstable. For $J_\perp J_z>0$, one finds ${\cal L}=-{\cal
H}+\gamma_2J_z(S_{az}+S_{bz})^{2}$ is minimal at the fixed point as
$\gamma_2\rightarrow\infty$. For $J_\perp J_z<0$ one finds ${\cal
L}={\cal H}+\gamma_2J_z(S_{az}+S_{bz})^{2}$, which is minimal at the
fixed point  as $\gamma_2\rightarrow\infty$. Therefore the fixed
point is stable when $J_\perp>0$, $J_\perp>-\eta_2J_z$ or
$J_\perp<0$, $J_\perp<-\eta_2J_z$.

(4) $\mathbf{n}_a =\mathbf{n}_b = (0,0,\pm  1)$; that is, the two spins are both parallel
or antiparallel to the $z$ direction. At each of these two fixed points,
the eigenvalues of ${\cal J}$ are $\mu_1=\mu_2=0$,
$\mu_{3,4}=\pm\sqrt{\frac{\zeta_{1}-\zeta_{2}}{2}},
\mu_{5,6}=\pm\sqrt{\frac{\zeta_{1}+\zeta_{2}}{2}}$; here $ \zeta_{1}
\equiv -2J_\perp^{2}S_{a}S_{b}-J_z^2(S_{a}^{2}+S_{b}^{2}),$ $
\zeta_{2} \equiv
J_z(S_{a}+S_{b})\sqrt{J_z^2(S_{a}-S_{b})^{2}+4J_\perp^{2}S_{a}S_{b}}
$. The stabilization cannot be judged by the eigenvalues. But one
finds the Lyapunov function ${\cal L}=-(S_{az}+S_{bz})^{2}$, which
is minimal at the fixed point. Hence these two fixed points are
always stable.

(5) In case  $J_\perp=J_z$, the solution
$\mathbf{n}_a=\mathbf{n}_{b}$ with any possible is a fixed point;
that is, the two spins are always parallel. The Lyapunov function
${\cal L}=-\mathbf{S}_a\cdot\mathbf{S}_{b}$ is minimal here, thus
this fixed point is stable.

(6) In case $J_\perp=J_z$, the solution
$\mathbf{n}_a=-\mathbf{n}_{b}$ with any possible is a fixed point;
that is, the two spins are always antiparallel. The Lyapunov
function ${\cal L}=\mathbf{S}_a\cdot\mathbf{S}_{b}$ is minimal here,
thus this fixed point is stable.

(7) In case $J_\perp=-J_z$, $\mathbf{n}_a=(\sin\theta\cos\varphi,
\sin\theta\sin\varphi,\cos\theta)$ while
$\mathbf{n}_b=(\sin\theta\cos\varphi,
\sin\theta\sin\varphi,-\cos\theta)$ is a fixed point; that is, the
$z$ components of the two spins are opposite.  One finds a Lyapunov
function ${\cal L}=-\mathbf{S}_a \cdot\mathbf{S}_b'$, where
$\mathbf{S}_b'=(S_{b}\sin\theta\cos\varphi,
S_{b}\sin\theta\sin\varphi,S_{b}\cos\theta)$,  which is minimal at
this  fixed point. Thus this fixed point is stable.

(8) In case $J_\perp=-J_z$, $\mathbf{n}_a=(\sin\theta\cos\varphi,
\sin\theta\sin\varphi,\cos\theta)$ while
$\mathbf{n}_b=(-\sin\theta\cos\varphi,
-\sin\theta\sin\varphi,\cos\theta)$ is a fixed point; that is, the
$x$ and $y$ components of the two spins are opposite. One finds a
Lyapunov function ${\cal L}=\mathbf{S}_a\cdot \mathbf{S}_b''$, where
$\mathbf{S}_b''=(-S_{b}\sin\theta\cos\varphi,
-S_{b}\sin\theta\sin\varphi,-S_{b}\cos\theta)$,  which is minimal at
this  fixed point.  Thus this fixed point is stable.

All of the fixed points and their stable regimes are
listed in Table \ref{table1}.

\end{document}